\documentclass[%
 reprint,
superscriptaddress,
 amsmath,amssymb,
 aps,
floatfix,
]{revtex4-1}

\usepackage{graphicx}
\usepackage{dcolumn}
\usepackage{bm}
\usepackage{amssymb,amsmath,amsfonts,braket}
\usepackage{hyperref}
\usepackage{color}
\usepackage[utf8]{inputenc}
\usepackage{csquotes}

\newcommand{\RomanNumeralCaps}[1]{\MakeUppercase{\romannumeral #1}}

\begin{document}


\title{Temperature dependence of the topological phase transition of BiTeI from
first principles}

\author{V\'eronique Brousseau-Couture}
\email{veronique.brousseau.couture@umontreal.ca}
\affiliation{%
D\'epartement de physique and Regroupement Qu\'eb\'ecois sur les Mat\'eriaux
de Pointe, Universit\'e de Montr\'eal, Montr\'eal, Qu\'ebec, Canada
}%

\author{Gabriel Antonius}
\affiliation{
D\'epartement de chimie, biochimie et physique, Institut de recherche sur
l'hydrog\`ene, Universit\'e du Qu\'ebec à Trois-Rivi\`eres, Trois-Rivi\`eres,
Qu\'ebec, Canada
}%

\author{Michel C\^ot\'e}%
\affiliation{%
D\'epartement de physique and Regroupement Qu\'eb\'ecois sur les Mat\'eriaux
de Pointe, Universit\'e de Montr\'eal, Montr\'eal, Qu\'ebec, Canada
}%

\date{\today}

\begin{abstract}

A topological phase transition from a trivial insulator to a $\mathbb{Z}_2$
topological insulator requires the bulk band gap to vanish. In the case of
noncentrosymmetric materials, these phases are separated by a gapless Weyl
semimetal phase. However, at finite temperature, the gap is affected by atomic
motion, through electron-phonon interaction, and by thermal expansion of the
lattice. As a consequence, the phase space of topologically nontrivial phases
is affected by temperature. In this paper, the pressure and temperature
dependence of the indirect band gap of BiTeI is investigated from first
principles. We evaluate the contribution from both electron-phonon interaction
and thermal expansion, and show that their combined effect drives the
topological phase transition towards higher pressures with increasing
temperature. Notably, we find that the sensitivity of both band extrema to
pressure and topology for electron-phonon interaction differs significantly
according to their leading orbital character. Our results indicate that the
Weyl semimetal phase width is increased by temperature, having almost
doubled by 100~K when compared to the static lattice results. Our findings
thus provide a guideline for experimental detection of the nontrivial phases
of BiTeI and illustrate how the phase space of the Weyl semimetal phase in
noncentrosymmetric materials can be significantly affected by temperature.



\end{abstract}

\maketitle

\section{\label{intro}Introduction}

Topological phases of matter have become a thriving field of condensed matter
physics, for both fundamental and applied research~\cite{hasan_colloquim_2010}.
In three-dimensional (3D) materials, these phases are characterized by the existence of metallic
surface states with peculiar properties, such as protection against
backscattering, spin-momentum locking and dissipationless spin-polarized
currents, as well as inverted bulk band
gaps~\cite{hasan_colloquim_2010,hasan_topological_2015}. The discovery of
experimentally tunable topological phases, whether through stoichiometric
doping~\cite{xu_topological_2011,wojek_band_2014,dziawa_topological_2012,
hsieh_topological_2008}, hydrostatic
pressure~\cite{xi_signatures_2013,bera_sharp_2013},
strain~\cite{li_robust_2016,liu_tuning_2014}, external electric
fields~\cite{liu_switching_2015,zhang_electric-field_2013} or interaction with
light~\cite{wang_observation_2013,hubener_creating_2017}, has led to a
continually growing number of proposals for promising and innovative
applications relying on the refined engineering of these robust states and
their associated phase
transitions~\cite{fleet_topological_2015,zhu_topological_2013,
zhou_engineering_2014,alicea_majorana_2010}.

Another widely studied class of materials is the bulk Rashba semiconductors,
in which a strong spin-orbit interaction combined with the absence of inversion
symmetry leads to a splitting of electronic bands of opposite spin
polarization~\cite{bychkov_properties_1984,manchon_new_2015}. The band extrema
are shifted away from the time-reversal invariant points in the Brillouin zone,
both in energy and momentum, in the plane perpendicular to the potential
gradient~\cite{bahramy_bulk_2017}. This Rashba effect gives rise to many
quantum phenomena, such as the Edelstein, spin Hall and spin galvanic effects,
as well as noncentrosymmetric superconductivity [see, for example, Bahramy and
Ogawa~\cite{bahramy_bulk_2017} and references therein]. Just like the
topological surface states, the bulk Rashba split bands have their spin
orientation locked perpendicular to their momentum, leading to a helical spin
texture~\cite{murakawa_detection_2013}. This feature makes these materials
promising candidates for realizing devices involving the active manipulation
of the spin degree of freedom, for both
spintronics~\cite{zutic_spintronics:_2004} and quantum
computing~\cite{das_sarma_spin_2001}.

Because it exhibits both of these phenomena, there is no wonder that BiTeI has
attracted such a wide interest for the last decade, both from the experimental
and first-principles communities. Besides displaying one of the largest Rashba
splittings known so far~\cite{ishikaza_giant_2011,crepaldi_giant_2012,
landolt_disentanglement_2012,sakano_strongly_2013,bahramy_origin_2011,
eremeev_ideal_2012}, it was predicted to turn into a strong $\mathbb{Z}_2$
topological insulator under hydrostatic pressure~\cite{bahramy_emergence_2012}.
Following this prediction, BiTeI has been widely investigated through
optical~\cite{xi_signatures_2013,tran_infrared-_2014,makhnev_optical_2014},
electrical transport~\cite{jin_superconductivity_2017,qi_topological_2017} and
Shubnikov-de Haas oscillations
experiments~\cite{ideue_pressure_2014,martin_quantum_2013,
murakawa_detection_2013,park_quantum_2015,vangennep_pressure_2014}. Many
experiments revealed signatures supporting the existence of the topological
phase transition (TPT)~\cite{xi_signatures_2013,jin_superconductivity_2017,
park_quantum_2015,ponosov_dynamics_2014,qi_topological_2017}.


While at first it was thought that the TPT of BiTeI occurred at a single
critical pressure $P_C$, 
it was later demonstrated that the lack of inversion symmetry imposed the
existence of an intermediate Weyl semimetal (WSM) phase~\cite{liu_weyl_2014}
between the trivial band insulator and topological insulator phases, yielding
\emph{two} critical pressures. Tight-binding~\cite{liu_weyl_2014} and
first-principles calculations~\cite{rusinov_pressure_2016,facio_strongly_2018}
predicted that this WSM phase could exist only within a narrow pressure range
of 0.1-0.2 GPa, making its experimental detection technically challenging.

The precise value of the critical pressure of BiTeI is still elusive to this
day. Experiments have located it between 2.0 and
3.5~GPa~\cite{xi_signatures_2013,jin_superconductivity_2017,
park_quantum_2015,qi_topological_2017,ideue_pressure_2014}, while
first-principles calculations have predicted it in a slightly wider range of
pressures, from 1.6 to 4.5~GPa~\cite{bahramy_emergence_2012,rusinov_pressure_2016,
facio_strongly_2018,guler-kilic_pressure_2016}. From a theoretical point of
view, predicting the critical pressure comes down to finding the gap closing
pressure, which is inherently dependent on the accuracy of the calculated band
gap at ambient pressure, which can be biased by the well-established
underestimation of the band gap by density functional
theory~(DFT)~\cite{martin_electronic_2004}.

On the other hand, one should not forget that most first-principles
calculations are done under the assumption of a static lattice, while
experiments are inherently done at finite temperature.
This completely overlooks the fact that electrons can interact with thermally
activated phonons, resulting in a temperature-dependent shift in the electronic
eigenenergies. Moreover, at ${T=0}$~K, it neglects the contribution of the
zero-point renormalization (ZPR) to the eigenenergies, due to the zero-point
motion of the ions.  For narrow gap materials, this renormalization could, in
principle, push the system towards a band inversion. In the case of BiTeI,
this shift will directly affect \emph{both} critical pressures at which the
band gap will effectively close or reopen at a given temperature.


The ability of electron-phonon interaction (EPI) to induce a change in the
bulk topology was first demonstrated with model
Hamiltonians~\cite{garate_phonon-induced_2013,saha_phonon-induced_2014,
li_conductivity_2013}, and later analyzed by first-principles calculations for
BiTl(S$_{1-x}$Se$_x$)$_2$~\cite{antonius_temperature-induced_2016} and for the
Bi$_2$Se$_3$ family~\cite{monserrat_temperature_2016}. These studies were able
to capture the whole complexity of the EPI, with fewer approximations. They
demonstrated that, depending on the strength and sign of the different types
of couplings, nontrivial topology could either be promoted or suppressed by
temperature in real materials. The suppression of the topological surface
state signatures at sufficiently high temperatures was observed experimentally
in Pb$_{1-x}$Sn$_x$Se~\cite{dziawa_topological_2012,reijnders_optical_2014,
wojek_direct_2015}. The temperature-dependent band structure of BiTeI was
previously investigated by focusing on the variation of the Rashba splitting
rather than on tracking the TPT~\cite{monserrat_temperature_2017}.

In this paper, we study the temperature dependence of the topological phase
transition in BiTeI using first-principles methods, assessing for both the EPI
and thermal expansion (TE) contributions. We explicitly assess the temperature
dependence of the band-gap renormalization as a function of pressure, both in
the trivial and the topological phases. We observe that the EPI contribution
changes sign as the system undergoes the TPT, and link this behavior to the
band inversion phenomenon, showing that the band extrema exhibit a distinct
pressure dependence related to their leading orbital character. In turn, the TE
contribution is unaffected by the leading orbital character of the band
extrema, but manifests sensitivity to the topological invariant by changing
sign as the system undergoes the TPT. We finally evaluate how the pressure
width of the WSM phase is affected by temperature by extrapolating the total
renormalization trends of each band towards the TPT, and find that the stronger
renormalization of the band gap in the TI phase widens the WSM with increasing
temperature.

The remainder of this paper is organized as follows. After presenting the
theoretical formalism used to obtain the EPI and TE contributions to the
electronic structure in Sect.~\ref{theory}, we summarize the computational
details in Sect.~\ref{computation}. Sect.~\ref{staticresults} presents our
analysis of the TPT and WSM phase for a static lattice at ${T=0}$~K, while
Sect.~\ref{tdepresults} and~\ref{WSMResults}
focus on the effects of temperature on both the band
gap and the TPT. We finally discuss the implications and limitations of our
results in Sect.~\ref{discussion} and summarize our findings.
\section{\label{theory}Methodology}
The temperature dependence of the band-gap energy, $E_g$, for a given constant
pressure, can be approximated at first order by the sum of two distinct
contributions~\cite{allen_temperature_1983}, namely the thermal expansion,
$\Delta E_g^{\text{TE}}$, and the renormalization of the electronic
eigenenergies by electron-phonon interactions at constant volume, $\Delta
E_g^{\text{EPI}}$ :
\begin{equation}
    \Delta E_g  \simeq \Delta E_g^{\text{EPI}} + \Delta E_g^{\text{TE}}.
\end{equation}

\subsection{Electron-phonon interaction}\label{AHC}
In many-body perturbation theory~\cite{mahan_many-particle_1990}, the
temperature dependence of the electronic eigenenergies,
$\varepsilon_{\mathbf{k}n}$, induced by electron-phonon interaction can be
captured, under certain assumptions~\cite{antonius_dynamical_2015}, by the real
part of the electron-phonon self-energy, $\Sigma^{\text{EPI}}$, evaluated at
the quasiparticle energy:
\begin{equation}
    \varepsilon_{\mathbf{k}n}(T) = \varepsilon^0 _{\mathbf{k}n} +
    \mathfrak{Re}\left[\Sigma^{\text{EPI}}_{\mathbf{k}n}(\omega=
    \varepsilon_{\mathbf{k}n}(T))\right],
\end{equation}
where $\varepsilon^0 _{\mathbf{k}n}$ is the unperturbed eigenvalue of a
Kohn-Sham (KS) eigenstate with wavevector $\mathbf{k}$ and band index $n$,
computed at the fixed, relaxed geometry. The Hartree atomic unit system is used
throughout this paper ($\hbar=m_e=e=1$).

\begin{figure}
    \centering
    \includegraphics[width=\linewidth]{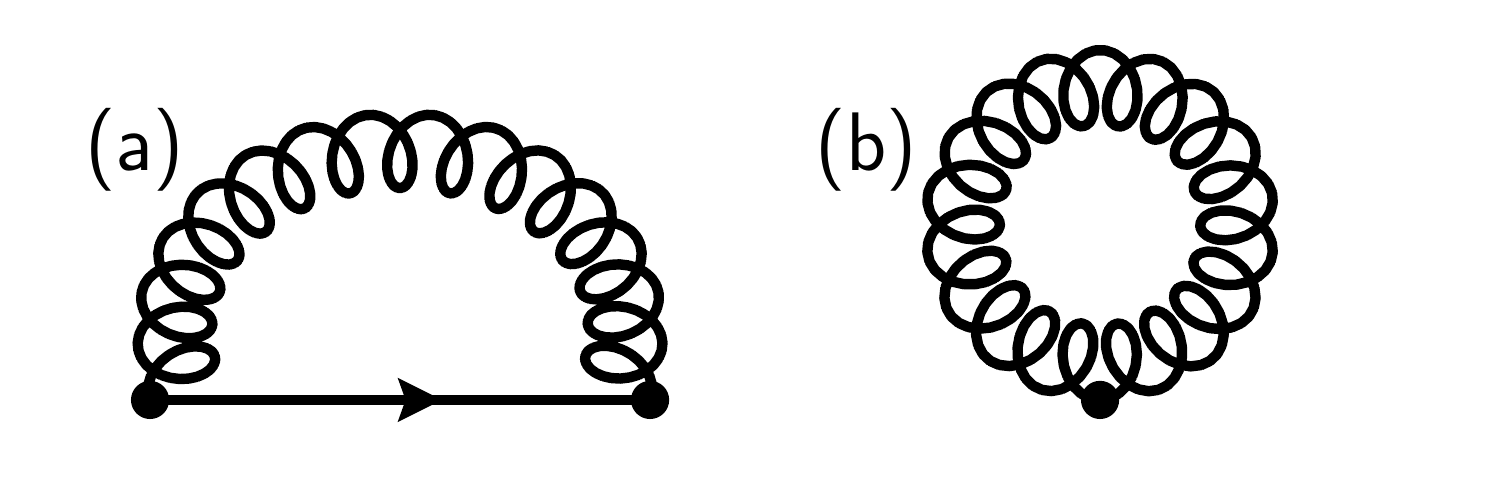}
    \caption{The Fan~(a) and Debye-Waller~(b) diagrams describing the lowest-order EPI in the AHC framework.}
    \label{FigFanDW}
\end{figure}

In the nonadiabatic Allen-Heine-Cardona (AHC)
framework~\cite{allen_theory_1976,allen_theory_1981,allen_temperature_1983},
the self-energy at the second-order of perturbation is the sum of the Fan and
Debye-Waller (DW) contributions [Fig.~\ref{FigFanDW}]:
\begin{equation}
    \Sigma^{\text{EPI}}_{\mathbf{k}n}(\omega,T) =
    \Sigma^{\text{Fan}}_{\mathbf{k}n}(\omega,T) +
    \Sigma^{\text{DW}}_{\mathbf{k}n}(T),
\end{equation}
where the former captures a frequency-dependent interaction with two
first-order EPI vertices, and the latter captures a static one with a single
second-order vertex. Evaluating the previous equation at absolute zero
temperature yields the ZPR, which arises from the zero-point motion of the
ions. For a detailed review of the AHC methodology, we refer our readers to
the works of Ponc\'e \emph{et
al.}~\cite{ponce_temperature_2014,ponce_temperature_2015} and to the extensive
review by Giustino~\cite{giustino_electron-phonon_2017}.

Assuming that the quasiparticle energy is close to the unperturbed electronic
energy, one can apply the on-the-mass-shell
approximation~\cite{marini_many-body_2015}, in which the Fan self-energy is
evaluated at the poles of the Green's function, namely at the frequency
$\omega$ corresponding to the bare eigenvalue $\varepsilon_{\mathbf{k}n}^0$,
yielding
\begin{equation}
    \varepsilon_{\mathbf{k}n}(T) \approx \varepsilon^0 _{\mathbf{k}n} +
    \mathfrak{Re}\left[\Sigma^{\text{EPI}}_{\mathbf{k}n}(\omega=
    \varepsilon^0_{\mathbf{k}n},T)\right].
\end{equation}
Furthermore approximating the fully interacting electronic wavefunction by the
noninteracting KS-DFT wavefunction, one obtains the standard result for the
retarded Fan self-energy~\cite{ponce_temperature_2014,
giustino_electron-phonon_2017}:
\begin{equation}\label{EqnSigmaFan}
\begin{split}
   & \Sigma_{\mathbf{k}n}^{\text{Fan}}(\omega,T) = \frac{1}{N_\mathbf{q}}
   \sum\limits_{\mathbf{q}\nu}^{\text{BZ}}\sum\limits_{n'}
   \frac{1}{2\omega_{\mathbf{q}\nu}}|\bra{\mathbf{k+q}n'}V^{(1)}_{\mathbf{q}
   \nu}\ket{\mathbf{k}n}|^2 \times \\
   & \left[ \frac{n_{\mathbf{q}\nu}(T) + 1 -
   f_{\mathbf{k+q}n'}(T)}{\omega-\varepsilon^0_{\mathbf{k+q}n'}-
   \omega_{\mathbf{q}\nu} + i\eta_\mathbf{k}} + \frac{n_{\mathbf{q}\nu}(T) +
   f_{\mathbf{k+q}n'}(T)}{\omega-\varepsilon^0_{\mathbf{k+q}n'}+
   \omega_{\mathbf{q}\nu} + i\eta_\mathbf{k}} \right],
    \end{split}
\end{equation}
in which the contributions from all phonon modes with frequency
$\omega_{\mathbf{q}\nu}$ are summed for all wavevectors $\mathbf{q}$ and
branches $\nu$ in the Brillouin zone (BZ). The whole temperature dependence of
this expression is captured by the bosonic and fermionic occupation factors,
respectively $n_{\mathbf{q}\nu}$ and $f_{\mathbf{k+q}n'}$. The parameter
$\eta_\mathbf{k}=\eta\,\text{sgn}(\varepsilon^0_{\mathbf{k}n}-\mu)$, where
$\eta$ is real and positive, preserves causality by correctly shifting the
poles of the Green's function in the complex plane.

Equation~(\ref{EqnSigmaFan}) implies the limit of an infinite number of phonon
wavevectors ($N_\mathbf{q}\rightarrow\infty$), leading to a vanishingly small
value of $\eta$.
$V^{(1)}_{\mathbf{q}\nu}$ is the first-order self-consistent change of the
local KS potential induced by the collective atomic displacement
$\textbf{R}^\nu_{\kappa\alpha}(\mathbf{q})$ along a given phonon mode
~\cite{antonius_dynamical_2015, ponce_temperature_2014,
giustino_electron-phonon_2017}:
\begin{equation}\label{EqnV1}
V^{(1)}_{\mathbf{q}\nu} =
\sum\limits_{\kappa\alpha}\frac{\partial V^{\text{KS}}}{\partial
\textbf{R}^\nu_{\kappa\alpha}(\mathbf{q})} = \sum\limits_{\kappa,\alpha}
V^{(1)}_{\kappa\alpha}(\mathbf{q}\nu),
\end{equation}
with
\begin{equation}\label{EqnDerivative}
\frac{\partial}{\partial \textbf{R}^\nu_{\kappa\alpha}(\mathbf{q})}=
U_{\nu,\kappa\alpha}(\mathbf{q})\sum\limits_l\text{e}^{i\mathbf{q}\cdot
\mathbf{R}_l}\frac{\partial}{\partial \mathbf{R}_{l\kappa\alpha}}.
\end{equation}
In these expressions, $\mathbf{R}_{l\kappa\alpha}$ denotes the displacement of
atom $\kappa$, located in unit cell $l$ with lattice vector $\mathbf{R}_l$,
in cartesian
direction $\alpha$; $\textbf{R}^\nu_{\kappa\alpha}(\mathbf{q})$ therefore
describes the collective atomic displacement of along the
$\mathbf{q}\nu$-phonon mode. $V^{(1)}_{\kappa\alpha}(\mathbf{q}\nu)$ refers to
the contribution of a displacement of atom $\kappa$ in direction $\alpha$ to
the full first-order potential $V^{(1)}_{\mathbf{q}\nu}$.
$U_{\nu,\kappa\alpha}(\mathbf{q})$ is the $\mathbf{q}\nu$-phonon displacement
vector, which is related to the phonon eigenvector,
$\xi^\nu_{\kappa\alpha}(\mathbf{q})$, through:
\begin{equation}
    \xi^\nu_{\kappa\alpha}(\mathbf{q}) = \sqrt{M_\kappa}
    U_{\nu,\kappa\alpha}(\mathbf{q}),
\end{equation}
with $M_\kappa$ being the mass of atom $\kappa$.

The static Debye-Waller self-energy is defined as the second derivative of the
potential with respect to two atomic displacements, evaluated at the first
order in perturbation theory~\cite{antonius_dynamical_2015} :
\begin{equation}
\Sigma^{\text{DW}}_{\mathbf{k}n} (T) = \frac{1}{N_\mathbf{q}}
\sum\limits_{\mathbf{q}\nu}\frac{1}{2\omega_{\mathbf{q}\nu}}\bra{\mathbf{k}n}
V^{(2)}_{\mathbf{q}\nu} \ket{\mathbf{k}n}\left[2n_{\mathbf{q}\nu}(T) +
1\right],
\end{equation}
where the second-order perturbation potential is
\begin{equation}
V^{(2)}_{\mathbf{q}\nu} = \frac{1}{2}\sum\limits_{\substack{\kappa\alpha \\
\kappa'\alpha'}}\frac{\partial^2 V^{\text{KS}}}{\partial
\textbf{R}^\nu_{\kappa'\alpha'}(\mathbf{-q})\partial
\textbf{R}^\nu_{\kappa\alpha}(\mathbf{q})}\;,
\end{equation}
with the derivatives defined as in Eq.~(\ref{EqnDerivative}).
Only the phonon occupation factor contributes to the temperature dependence of
this term since it does not involve any intermediate electronic state [as can
be seen from Fig.~\ref{FigFanDW}(b)]. The constant term inside the brackets
accounts for the ZPR contribution.

The numerical evaluation of this second-order derivative is a computational
bottleneck of density-functional perturbation theory (DFPT)~\cite{ponce_temperature_2015}. It can, however, be
circumvented by applying the rigid-ion approximation (RIA), which supposes that
the potentials created by each nucleus are independent of each other. Within
this approximation, thanks to the translational invariance of the lattice,
$V^{(2)}_{\mathbf{q}\nu}$ can be expressed in terms of the first-order
derivatives entering the Fan self-energy~\cite{ponce_temperature_2014} [see
Eq.~(\ref{EqnV1})]. The consequences of this approximation on the ZPR of
diamond have been discussed by Ponc\'e \emph{et
al.}~\cite{ponce_temperature_2014}. In this framework, the Debye-Waller
self-energy becomes~\cite{ponce_temperature_2014,
antonius_temperature-induced_2016}:

\begin{equation}\label{EqnSigmaDW}
\begin{split}
    \Sigma^{\text{DW,RIA}}_{\mathbf{k}n} =
    &\frac{1}{N_\mathbf{q}}\sum\limits_{\mathbf{q}\nu}^{\text{BZ}}
    \sum\limits_{n'}-\frac{1}{2\omega_{\mathbf{q}\nu}}
    \frac{|g^{\text{DW}}_{\mathbf{k}nn'}(\mathbf{q}\nu)|^2}
    {\varepsilon^0_{\mathbf{k}n}-\varepsilon^0_{\mathbf{k}n'}+i\eta}\\
    &\times \left[n_{\mathbf{q}\nu}(T) + \frac{1}{2}\right],
    \end{split}
\end{equation}
with
\begin{equation}
    \begin{split}
        &|g^{\text{DW}}_{\mathbf{k}nn'}(\mathbf{q}\nu)|^2 =\\
       & \qquad \quad \sum\limits_{\substack{\kappa\alpha \\ \kappa'\alpha'}}
        \left[U_{\nu,\kappa\alpha}(\mathbf{q})U^*_{\nu,\kappa\alpha'}
        (\mathbf{q}) +
        U_{\nu,\kappa'\alpha}(\mathbf{q})U^*_{\nu,\kappa'\alpha'}
        (\mathbf{q})\right] \\
       &\qquad \quad \times  \bra{\mathbf{k}n}
       V^{\text{DW}*}_{\kappa\alpha}\ket{\mathbf{k+q}n'} \bra{\mathbf{k+q}n'}
       V^{\text{DW}}_{\kappa'\alpha'}\ket{\mathbf{k}n},
    \end{split}
\end{equation}
and
\begin{equation}
  V^{\text{DW}}_{\kappa\alpha} = V^{(1)}_{\kappa\alpha}(\mathbf{q}=0,\nu).
\end{equation}

We finally evaluate the sum on band index $n'$ in Eq.~(\ref{EqnSigmaFan})
and~(\ref{EqnSigmaDW}) using the semi-static approach described in Ponc\'e
\emph{et al.}~\cite{ponce_temperature_2014} and Antonius \emph{et
al.}~\cite{antonius_dynamical_2015}. We compute the full, nonadiabatic
contribution for all bands up to a band index $M$, which we choose to be
well-separated in energy (i.e., more than 10 eV) from the first conduction
band. For the high-energy bands with band index $n'>M$, the phonon frequencies
in the denominators of Eq.~(\ref{EqnSigmaFan}) and~(\ref{EqnSigmaDW}) are
negligible with respect to the electronic eigenenergy difference. These bands
can, therefore, be treated within the adiabatic approximation, neglecting the
phonon frequencies. Furthermore, this sum over an arbitrarily large number of
empty states can be replaced by the solution of a linear Sternheimer equation
for the subspace orthonormal to the low-energy states ($n'\leq M$), thus
significantly reducing the numerical cost of the
calculation~\cite{gonze_theoretical_2011}. We estimate the relative error on
the renormalization induced by this semi-static treatment of the high energy
bands to be 1-2\% at most.


\subsection{Thermal expansion}\label{theoryTE}
The thermal expansion contribution was evaluated through the quasi-harmonic
approximation (QHA)~\cite{grimvall_thermophysical_1986,
srivastava_physics_1990}. Within this framework, the only departure from
harmonicity occurs through the explicit volume dependence acquired by the
phonon frequencies. It should be understood that the purpose of the QHA is to
deliver the leading order expression of the thermal expansion coefficient
$\alpha$, rather than capturing the full anharmonic effects present in the
crystal~\cite{allen_quasi-harmonic_2019}, as would do nonperturbative
methods~\cite{antonius_dynamical_2015}.

The thermal expansion coefficient, $\alpha$, is a rank~2 tensor that relates
a small temperature increment $\Delta T$ to the strain, $\epsilon$, it induces
on the lattice:
\begin{equation}
    \epsilon_{ij} = \alpha_{ij}\Delta T,
\end{equation}
where $i,j=1,2,3$ are cartesian directions.
Hence, the most general definition of $\alpha_{ij}$ is
\begin{equation}
    \alpha_{ij} = \left(\frac{\partial\epsilon_{ij}}{\partial T}\right)_\sigma,
\end{equation}
where the derivative is evaluated at constant stress $\sigma$, which is
usually taken to be a constant pressure $P$. The  diagonal components of the
strain tensor describe the relative change in the length of the lattice
parameters: $ \epsilon_{ii}=\Delta a_i/a_i$. The volumic expansion coefficient
$\beta$ can therefore be obtained by taking the trace of $\alpha$:
\begin{equation}
    \beta = \frac{\partial}{\partial T}\left(\frac{\Delta V}{V}\right)_P =
    \sum\limits_{i=1}^3 \alpha_{ii}.
\end{equation}

In the most simple case of cubic symmetry, all $\alpha_{ii}$ are equal and can
be expressed by~\cite{grimvall_thermophysical_1986}
\begin{equation}\label{EqnAlphaCubic}
    \alpha(T) = \frac{\beta}{3} = \frac{\kappa_0}
    {3V_0 N_{\mathbf{q}}}\sum\limits_{\mathbf{q}\nu}\gamma^V_{\mathbf{q}\nu}
    c_{\mathbf{q}\nu}(T),
\end{equation}
where $\kappa_0$ is the bulk compressibility at equilibrium volume, $V_0$ is the
primitive cell equilibrium volume,  $N_{\mathbf{q}}$ is the number of phonons in
the homogenous grid used to sample the BZ and $c_{\mathbf{q}\nu}(T)$ is the
specific heat of the $\mathbf{q}\nu$-phonon mode at temperature $T$. We have
also introduced the volumic mode Gr\"uneisen parameters, defined
as~\cite{grimvall_thermophysical_1986}
\begin{equation}
    \gamma^V_{\mathbf{q}\nu} =
    -\frac{\partial\text{ln}\omega_{\mathbf{q}\nu}(V)}{\partial\text{ln}V}.
\end{equation}
The bulk Gr\"uneisen parameter, often referred to in the literature, is
defined as
\begin{equation}\label{EqnBulkGru}
    \gamma^V = \frac{\sum\limits_{\mathbf{q}\nu}\gamma^V_{\mathbf{q}\nu}
    c_{\mathbf{q}\nu}}{\sum\limits_{\mathbf{q}\nu}c_{\mathbf{q}\nu}}.
\end{equation}
Thus, for cubic systems, the linear thermal expansion coefficient,
$\alpha(T)$, Eq.~(\ref{EqnAlphaCubic}), is simply proportional to two scalar
parameters, $\gamma^V$ and $\kappa_0$, the former governing the sign of the
thermal expansion.

For materials belonging to noncubic symmetry groups, one must consider the
most general case, in which the mode Gr\"uneisen parameters take a tensor form
and capture the variation of the phonon frequencies with respect to a given
strain:
\begin{equation}\label{EqnGammaij}
    \gamma^{ij}_{\mathbf{q}\nu} =
    -\frac{\partial\text{ln}\omega_{\mathbf{q}\nu}}{\partial\epsilon_{ij}}.
\end{equation}
One can also define bulk Gr\"uneisen parameters $\gamma^{ij}$ following the
same procedure as in Eq.~(\ref{EqnBulkGru}).

For axial crystals, which include the case of BiTeI's trigonal symmetry,
the thermal expansion coefficient tensor has two distinct nonvanishing
components$:\alpha_{11}=\alpha_{22}$ and $\alpha_{33}$.
Because of this anisotropy, the resulting linear thermal expansion
coefficients along crystallographic directions $\hat{a}$ and $\hat{c}$
capture a 
subtle interplay between the vibrational and elastic properties of the
material~\cite{munn_role_1972,ritz_interplay_2018,
grimvall_thermophysical_1986}:
\begin{equation}\label{EqnAlpha}
\begin{split}
    \alpha_a &= \alpha_{11} =
    \frac{C_\sigma}{V_0}\left[(s_{11}+s_{12})\gamma^{a_1} +
    s_{13}\gamma^c\right],\\
    \alpha_c &= \alpha_{33} = \frac{C_\sigma}{V_0}\left[2s_{13}\gamma^{a_1} +
    s_{33}\gamma^c\right],
    \end{split}
\end{equation}
where $C_\sigma$ is the heat capacity at constant stress and $s_{ij}$ is the
$ij$-coefficient of the elastic compliance tensor. In these expressions, the
subscripts $a$ and $c$ refer to the length of the unit cell along
$(\hat{a}_1,\hat{a}_2)$ and $\hat{c}$. $\gamma^{a_1}$ formally refers to a
strain applied in only \emph{one} direction perpendicular in the
crystallographic axis.

From Eq.~(\ref{EqnGammaij}), the directional mode Gr\"uneisen parameters
$\gamma^{a_1}_{\mathbf{q}\nu}$ and $\gamma^c_{\mathbf{q}\nu}$ take the form
\begin{equation}\label{EqnGammaHex}
    \begin{split}
        \gamma^{a_1}_{\mathbf{q}\nu} &= \gamma^{11}_{\mathbf{q}\nu} =
        -\frac{1}{2}
        \left(\frac{\partial\text{ln}\omega_{\mathbf{q}\nu}}
        {\partial\text{ln}a}\right)_c\;,\\
        \gamma^{c}_{\mathbf{q}\nu} &= \gamma^{33}_{\mathbf{q}\nu} = -
        \left(\frac{\partial\text{ln}\omega_{\mathbf{q}\nu}}
        {\partial\text{ln}c}\right)_a\;.\\
    \end{split}
\end{equation}
In the previous expressions, the phonon frequency derivatives are evaluated
with respect to the variation of only one lattice parameter, the other one
remaining fixed at its static equilibrium value. One should finally note that,
in contrast to Eq.~(\ref{EqnGammaij}), the derivative entering
$\gamma_{\mathbf{q}\nu}^{a_1}$ is taken by changing the $a$ cell dimension,
which affects \emph{both} lattice vectors in the basal plane, hence the factor
1/2. We finally note that the derivation of the Gr\"uneisen
formalism~\cite{grimvall_thermophysical_1986} neglects the zero-point energy
of the phonons, such that there is no thermal expansion at ${T=0}$~K.


\section{\label{computation}Computational details}
\subsection{First-principles calculations}
All first-principles calculations were performed using the ABINIT software
package~\cite{gonze_abinit_2019}. The bulk electronic structure was obtained
within DFT,
while response function and electron-phonon coupling calculations were
performed within DFPT~\cite{gonze_first-principles_1997,
gonze_dynamical_1997}. We used a
maximum
plane-wave energy of 35 hartree and sampled the BZ using a ${8\times8\times}8$
Monkhorst-Pack $\mathbf{k}$-point grid. Spin-orbit coupling was taken into
account as it is necessary to obtain the Rashba effect and since it has been
shown to strongly affect both electronic~\cite{bahramy_origin_2011} and
vibrational~\cite{sklyadneva_lattice_2012} properties of BiTeI. We used
Hartwigsen-Goedecker-Hutter (HGH) fully relativistic norm-conserving
pseudopotentials~\cite{hartwigsen_relativistic_1998}, including explicitly the
semi-core $5d$ electrons for Bi. The electron-phonon self-energy was computed
with the ElectronPhononCoupling module.

Throughout this paper, we rely on the generalized gradient approximation of the
Perdew-Burke-Ernzerhof (PBE-GGA) functional~\cite{perdew_generalized_1996},
although it has been shown to overestimate the lattice parameters at ambient
pressure for this particular
material~\cite{guler-kilic_crystal_2015,rusinov_pressure_2016,
monserrat_temperature_2017}. Since the purpose of this paper is to investigate
the temperature dependence of the TPT, we chose the functional that gives us
the best overall agreement with experiment for the bare band gap and the
projected orbital character of the band extrema at ambient pressure, as well
as for the predicted critical pressure $P_{\text{C1}}$. For detailed results
and a complete discussion about our choice of PBE-GGA functional, see Appendix~\ref{AppendixStructure}.

\subsection{Structural properties}

The unit-cell geometry 
has been optimized until the resulting forces on all atoms were lower than
$10^{-5}$~hartree/bohr$^{3}$. Due to the layered nature of BiTeI, the
in-plane
and normal lattice parameters of BiTeI do not vary isotropically under
hydrostatic pressure~\cite{xi_signatures_2013}: $c$ decreases more rapidly
than $a$ within the first GPa applied. In order to allow for this
nonhomogenous variation of the lattice parameters, the external pressure was
modelized by applying an isotropic stress tensor.
The lattice was fully optimized for nine different pressures between 0~and 5~GPa.
The resulting cell volumes were then fitted with a Murnaghan equation of
state (EOS)~\cite{murnaghan_compressibility_1944} in order to validate the
theoretically applied pressure. For all pressures below 3.5~GPa, the
discrepancy between the EOS pressure and the applied pressure was less than
0.1~GPa, and less than 0.06~GPa for pressures surrounding the TPT.
The lattice structure at intermediate pressures was obtained by interpolating
from the results of those nine calculations.
The optimized lattice structure at all
pressures considered in this paper can be found in
Table~\ref{TableOpt} of Appendix~\ref{AppendixStructure}.

%


\subsection{Lattice dynamics and electron-phonon coupling}

Phonon frequencies and electron-phonon matrix elements were computed with a
$12\times12\times12$ homogeneous, $\Gamma$-centered Monkhorst-Pack
$\mathbf{q}$-point grid. Explicit calculations were done at seven different
pressures, spanning both the trivial (0.0, 0.5, 1.0, and 1.5~GPa) and
topological (3.0, 3.5~and 5.0~GPa) phases. The pressure dependence of the
phonon frequencies for the Raman-active modes is in good agreement with
experimental data [see Tables~\ref{TablePhononfreqZ0}~and~\ref{TablePhononfreqZ1} of Appendix~\ref{AppendixStructure}].

In a typical nonadiabatic AHC calculation, one aims for a value of $\eta$
smaller than the highest phonon energy (typically around $\eta=0.01$ eV, see,
for example, Nery \emph{et al.}~\cite{nery_quasiparticles_2018}). 
This however requires a very dense $\mathbf{q}$-point sampling. As discussed in
the Appendix of Antonius \emph{et al.}~\cite{antonius_dynamical_2015}, when
using a coarser $\mathbf{q}$-point grid, one risks artificially emphasizing
the contribution of a discrete number of terms in Eq.~(\ref{EqnSigmaFan})
where, for a given ${\mathbf{q}\nu}$-phonon, the value of
$\varepsilon^0_{\mathbf{k}n}-\varepsilon^0_{\mathbf{k+q}n'}
\pm\omega_{\mathbf{q}\nu}$ gets vanishingly small, hence inducing rapid
fluctuations of the self-energy~\cite{antonius_dynamical_2015}. These peaks are
a numerical artefact of the $\mathbf{q}$-point sampling, and the presence of
such fluctuations in the vicinity of the bare electronic eigenvalue
$\varepsilon^0_{\mathbf{k}n}$ could lead to an overestimation of the
renormalization. In contrast, increasing $\eta$ without caution can lead to
an unphysical flattening of the self-energy, suppressing physically relevant
features. We therefore chose $\eta$=0.1 eV, which allows the self-energy to be
a smooth function of frequency, without suppressing physically relevant
features. We verified that this choice was suitable for the claims reported in
this paper.

Lastly, our first-principles methodology relies on the assumption that the
electron-phonon interaction can modify the electronic eigenenergies, but cannot
by itself promote an electron to an excited state. Such a possibility would
modify the resulting electronic density and would require one to evaluate the
nondiagonal elements of the self-energy, $\Sigma_{\mathbf{k}n,n'}$ [see Eq.
(4) of Antonius \emph{et al.}~\cite{antonius_dynamical_2015}], allowing
$n'\neq n$. This procedure was, however, out of reach for the present work.
Thus, we restricted our investigation of electron-phonon coupling to pressures
where the predicted bare band gap was greater than the highest phonon energy
(E$_g>20$~meV).


\subsection{Thermal expansion}
The directional Gr\"uneisen parameters $\gamma^{a_1}$ and $\gamma^c$ were
evaluated  using an $8\times8\times8$ $\Gamma$-centered Monkhorst-Pack
\mbox{$\mathbf{q}$-point} grid.
Note that the Gr\"uneisen parameters require a smaller $\mathbf{q}$-point grid
than the EPI interaction since they only depend on the phonon frequencies,
which converge faster with the $\mathbf{q}$-point sampling.

The derivatives in Eq.~(\ref{EqnGammaHex}) were computed by central finite
difference using three different volumes per direction, per pressure. For each
calculation, the internal atomic coordinates in the $\hat{c}$ direction were
optimized at fixed volume, using the same convergence criterion as for the
equilibrium structure. The compliance constants $s_{ij}$ were obtained by
computing the strain-strain derivatives of the total energy within
DFPT~\cite{hamann_metric_2005}. We used the so-called relaxed-ion compliance
tensor, which takes into account the relaxation of the atomic coordinates
within the unit cell. This typically lowers the components of the elastic
stiffness tensor (usually referred to as the elastic constants), as the
resulting stress on the atoms is reduced by the relaxation process. Since the
compliance tensor is the inverse of the elastic stiffness tensor, it typically
increases the elastic compliance when compared to the clamped-ion
result~\cite{wu_systematic_2005}.


\section{Results and discussion}

\subsection{Topological phase transition in the static
lattice}\label{staticresults}
\begin{figure}
    \centering
        \includegraphics[width=0.5\linewidth]{fig2a_crystal_col-eps-converted-to.pdf}%
\includegraphics[width=0.5\linewidth,trim={0 0 0 0},clip]{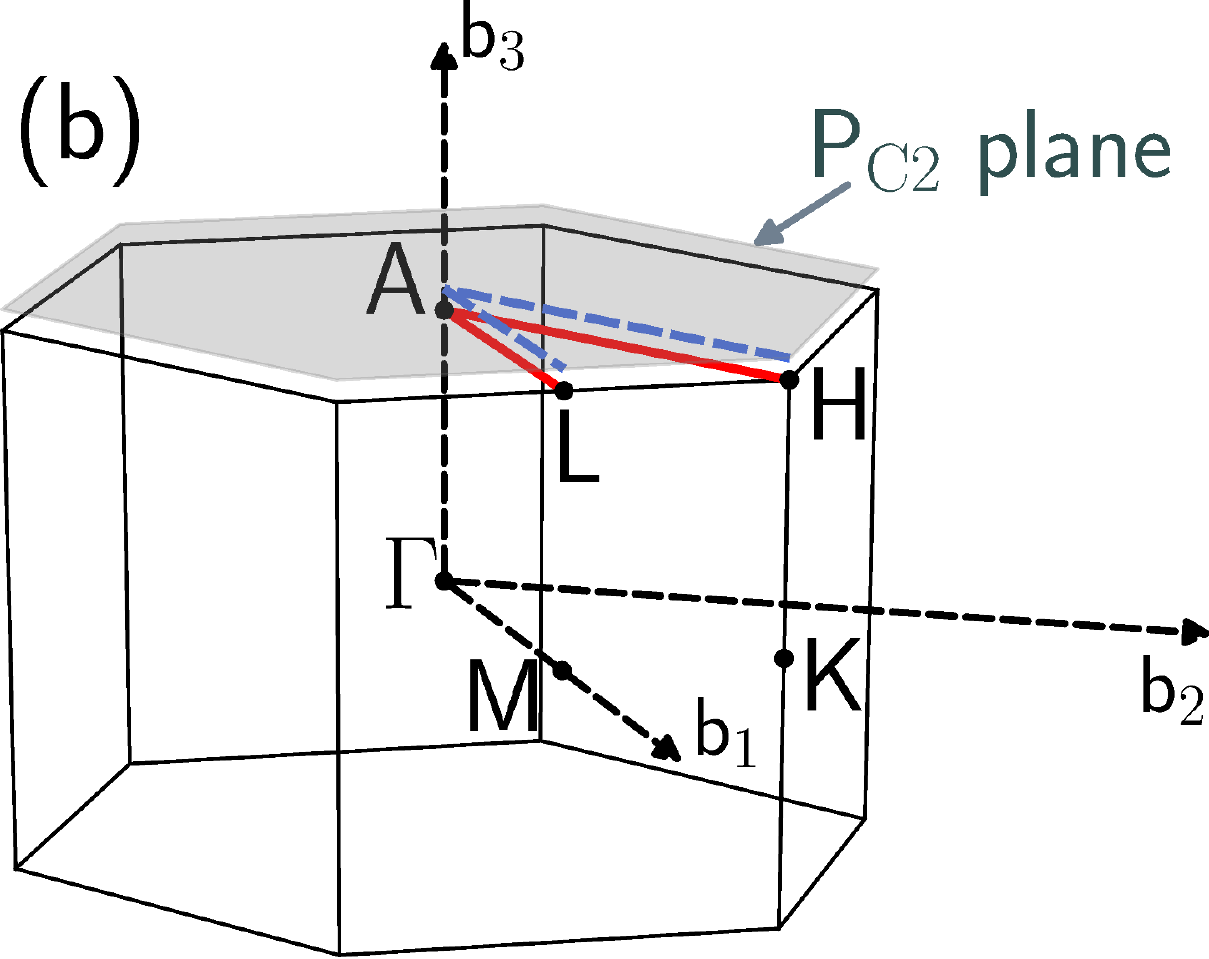}
    \caption{\textbf{Crystal structure and first Brillouin zone of BiTeI.}
    (a)~The different I-Bi-Te trilayers are weakly bound by van der Waals
    interaction along the crystallographic $c$ axis. Dashed lines delimit a
    unit cell. (b)~The path used for static band structure, EPI, and TE
    calculations in the trivial insulator phase is shown in solid red. The
    shaded gray area corresponds to the P$_{\text{C2}}$ plane (defined in
    Section~\ref{staticresults}), where the Weyl points annihilate and the
    system enters the TI phase. For $P>P_{\text{C2}}$, the band structure was
    computed along the dashed blue path.}\label{FigStructure}
\end{figure}

\begin{figure*}
\centering
\includegraphics[width=\linewidth,trim={0.5cm 0 0.5cm 0},clip]{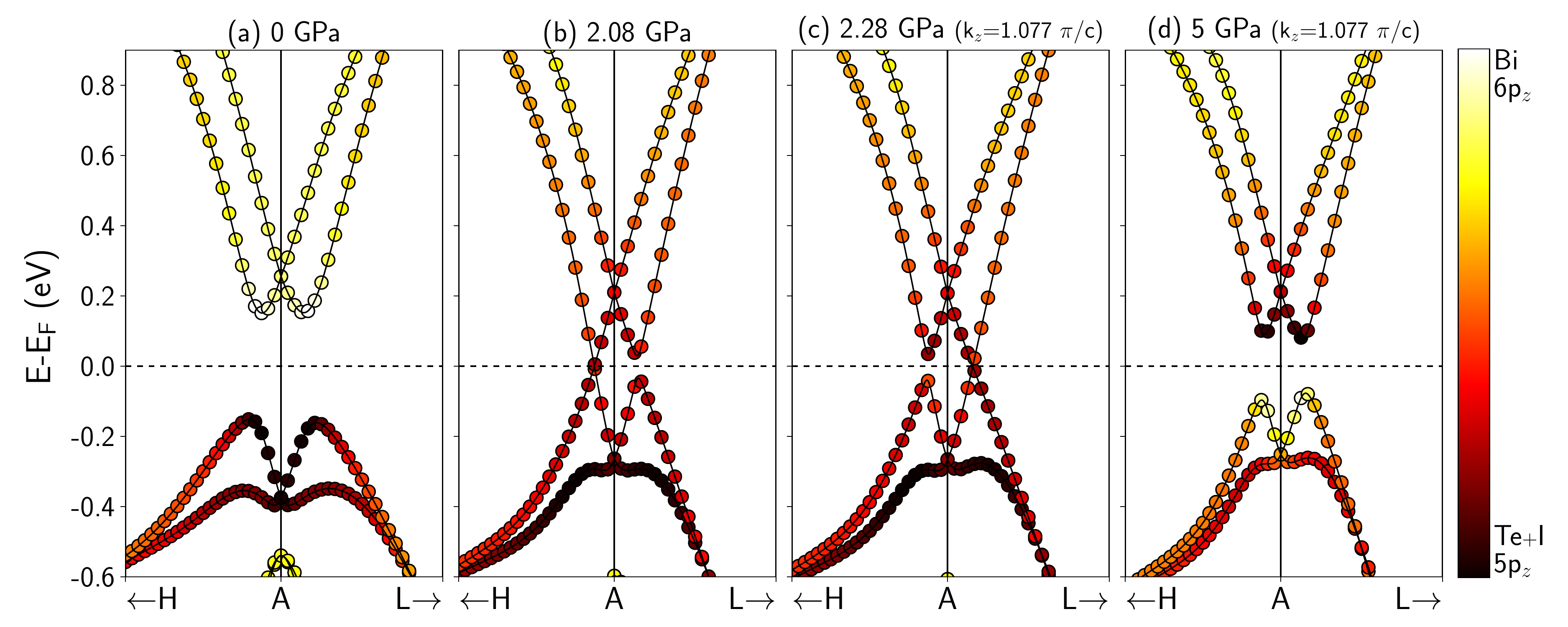}
\caption{\textbf{Band inversion process and critical pressures.} The solid black
lines show the bulk band structure at a given pressure, while the colored markers
describe the relative character of the projection of the wavefunction on the
$l=1,m=0$ spherical harmonic centered around the different atoms in the unit
cell. It displays the normalized difference between Bi-6p$_z$ and
(Te+I)-5p$_z$ projections. A band with leading Bi character will show in light yellow, while dark brown 
refers to a leading Te+I character. The band inversion of
the Bi and Te/I states observed at 5~GPa (d), when compared to ambient pressure (a),
is a signature of the TPT. At the two critical pressures $P_{\text{C1}}$ =
2.08~GPa (b)  and $P_{\text{C2}}$ = 2.28~GPa (c), the
band structure displays the cross-section of a Dirac cone, along H-A  for the
former and in the A-L-M mirror plane for the latter. For all panels, the
horizontal axis shows a path representing $\sim 15\%$ of the H-A and A-L
distances. For $P\leq P_{C1}$, the displayed band structure is located at
$k_z=\pi/c$, while for $P\geq P_{C2}$ it is shifted to the $P_{\text{C2}}$
plane [shaded area and dashed blue path of
Fig.~\ref{FigStructure}(b)].}\label{FigFatbands}
\end{figure*}
\begin{figure}
\centering
\includegraphics[width=\linewidth]{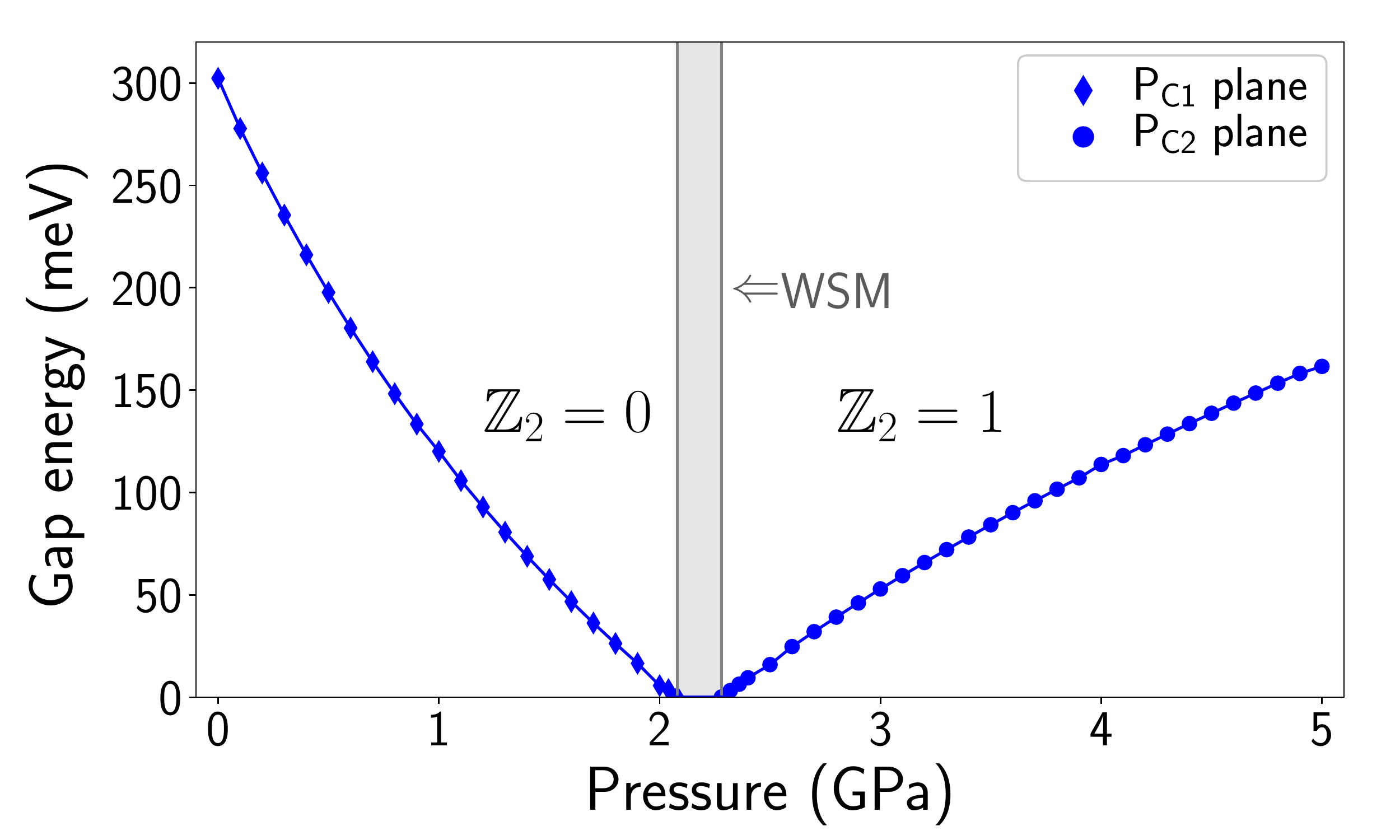}
\caption{\textbf{Variation of the indirect band gap with pressure.} In the
trivial insulator phase ($\mathbb{Z}_2=0$), the minimal band gap is located in
the H-A-L plane (diamonds), where the first band touching point will occur at
$P_{\text{C1}}$=2.08~GPa. The system enters the topological phase
($\mathbb{Z}_2=1$) once the pairs of Weyl points have annihilated at the
second band touching point, at $P_{\text{C2}}$=2.28~GPa. For higher pressures,
the minimal band gap is now located in the A-L-M mirror plane, with
$k_z=1.07698\;\pi/c$ (circles). The gray shaded area represents the WSM phase
width in the static lattice approximation. See Section~\ref{staticresults}
for more details.}\label{FigBareGapP}
\end{figure}

BiTeI is a layered material composed of alternating Bi, Te and I planes
stacked along the high-symmetry crystallographic $\hat{c}$ axis. This small
band-gap semiconductor belongs to the noncentrosymmetric trigonal space group
P3m1 (no. 156, mp-22965). The crystal structure and first BZ are shown in
Fig.~\ref{FigStructure}.

A common signature of a TPT is an inversion of the orbital character of the
band structure in the vicinity of the band gap~\cite{hasan_colloquim_2010}. In
the case of BiTeI, a schematic analysis of the band splitting showed that the
inverted bands should be of $p_z$ character~\cite{bahramy_emergence_2012}.
Fig.~\ref{FigFatbands} displays the leading orbital character of the last
valence and first conduction bands, for different pressures throughout the
studied range. We computed the $p_z$ character by projecting the wavefunction
on the $l=1, m=0$ spherical harmonic centered around the different atoms in
the unit cell. To emphasize the band inversion, Fig.~\ref{FigFatbands}
shows the \emph{relative} projected character on each band, namely the
normalized difference between the Bi-6p$_z$ and Te/I-5p$_z$ projections.

At ambient pressure [Fig.~\ref{FigFatbands}(a)], the indirect gap
is located
in the $k_z=\pi/c$ plane, close to the A high-symmetry point, in the
direction of H [on the red path of Fig.~\ref{FigStructure}(b)].  In the H-A and
A-L directions, which both exhibit Rashba splitting, the valence band maximum
(VBM) is dominated by $p_z$ states of Te and I (dark brown), whereas the conduction
band minimum (CBM) has Bi-$p_z$ (light yellow) character.  It is a trivial band
insulator, with topological index $\mathbb{Z}_2$=(0;000). The $\mathbb{Z}_2$
topological index was computed by tracking the evolution of the hybrid Wannier
charge centers, using the Z2Pack software
package~\cite{gresch_z2pack:_2017,soluyanov_computing_2011}. For more
information about the definition of the $\mathbb{Z}_2$ topological index for
3D topological insulators, we refer to Sections \RomanNumeralCaps{2}-C and
\RomanNumeralCaps{4}-A of the review by Hasan and
Kane~\cite{hasan_colloquim_2010}.

As pressure is increased, the band gap progressively decreases until it closes
at $P_{\text{C1}}$=2.08~GPa~[Fig.~\ref{FigFatbands}(b)]. At this
point, the BZ
exhibits three pairs of Dirac cones, which
are split into pairs of Weyl nodes upon further pressure increase; see Fig.~11
of Liu and Vanderbilt~\cite{liu_weyl_2014}. These Weyl nodes migrate within
the BZ until they annihilate each other,
at $P_{\text{C2}}$=2.28~GPa~[Fig.~\ref{FigFatbands}(c)].

Fortunately, from previous studies~\cite{liu_weyl_2014,rusinov_pressure_2016},
it is known that the gap closing occurs along symmetry line H-A (with
$k_z=\pi/c$), and that it will reopen along a symmetry line rotated by
$\pi/6$, without symmetry constraints in the $k_z$ direction (namely, in the
A-L-M mirror plane). In the following, we define the $P_{\text{C}2}$ plane
as the plane parallel to the H-A-L plane, with $k_z=1.07698\;\pi/c$, which is
the $k_z$ coordinate of the second band touching point displayed in
Fig.~\ref{FigFatbands}(c), where the Weyl nodes' annihilation
occurred. This plane corresponds to the shaded plane in
Fig.~\ref{FigStructure}(b). In a similar fashion, we identify the H-A-L plane
as the $P_{\text{C}1}$ plane.

In the vicinity of the TPT, the valence and conduction band extrema show a
mixed character, with an almost equal contribution of Bi-p$_z$ and Te/I-p$_z$
states~[Fig.\ref{FigFatbands}(b-c)], foreshadowing the
band
inversion [Fig.~\ref{FigFatbands}(d)] which occurs for $P>$2.28~GPa.
The
system is then a strong topological insulator, with $\mathbb{Z}_2$=(1;001).
In the TI phase, the minimal band gap is shifted from the H-A-L plane
($k_z=\pi/c$) to the $P_{\text{C2}}$ plane ($k_z=1.07698\;\pi/c$),
where the Weyl nodes' annihilation occurred. The full pressure dependence of
the band gap for the static lattice is shown in Fig.~\ref{FigBareGapP}.


For the static lattice at ${T=0}$~K, we thus obtain a WSM phase width of
0.2~GPa,
in good agreement with previous
calculations~\cite{rusinov_pressure_2016,facio_strongly_2018,liu_weyl_2014}.
Our value for the first gap closing $P_{\text{C1}}$=2.08~GPa agrees with many
experimental estimations~\cite{jin_superconductivity_2017,
xi_signatures_2013,park_quantum_2015} but is slightly lower than those of
other experimental~\cite{ideue_pressure_2014,qi_topological_2017} and
computational works~\cite{rusinov_pressure_2016,chen_high-pressure_2013,
facio_strongly_2018}. For comparison with experiments, one should note that,
as the TPT of
BiTeI occurs under hydrostatic pressure, direct observation of the
topological surface states with angle-resolved photoemission spectroscopy (ARPES) would be challenging from an
experimental point of view. As a consequence, experiments rather focused on
indirect signatures of these states, for example, a broad minimum of
resistivity~\cite{jin_superconductivity_2017,qi_topological_2017}, which
do not allow a precise estimation of $P_C$. The difference with other
computational works can be explained by the use of experimental
pressure-dependent lattice parameters instead of the
DFT-relaxed ones~\cite{facio_strongly_2018}, as well as by the
choice of projector-augmented wave (PAW) pseudopotentials without semi-core
states for Bi~\cite{rusinov_pressure_2016,chen_high-pressure_2013} and the use
of the Heyd-Scuseria-Ernzerhof (HSE) hybrid
functional~\cite{chen_high-pressure_2013} to compute the static gap at ambient
pressure.
Indeed, the calculated critical pressure is quite sensitive to the choice of
functional, as it may significantly modify the starting gap at ambient
pressure. One must also recall that the DFT gap has fundamentally no physical
meaning; a proper, theoretically adequate calculation of the ambient pressure
gap would require using a many-body methodology like GW.
Nevertheless, for the scope of this paper, we are more interested in the
\emph{relative} variation of the critical pressures rather than in their
absolute values. These slight discrepancies with other works do not,
therefore, undermine our conclusions.


Our value of $P_{\text{C1}}$ corresponds to a relative volume ($V$/$V_0$)
compression of $\sim12\%$, in good agreement with other calculations relying
on the PBE functional~\cite{bahramy_emergence_2012}. We also verified that the
relative volume compression within the WSM phase was consistent with the
original prediction of Liu and Vanderbilt~\cite{liu_weyl_2014}.

\subsection{Temperature dependence of the topological phase
transition}\label{tdepresults}

Within this paper, we track the temperature dependence of the critical
pressures $P_{\text{C1}}$ and $P_{\text{C2}}$ by evaluating how the
electron-phonon interaction and the thermal expansion will affect the indirect
band-gap value. Our analysis is based on several assumptions. First, we
suppose that the $\mathbb{Z}_2$ topological index computed at ${T=0}$~K for the
static lattice remains unchanged unless the renormalization closes the gap.
Second, since the system lacks inversion symmetry regardless of temperature,
we assume that Liu and Vanderbilt's argument~\cite{liu_weyl_2014} is still
valid, such that a gapless WSM phase still separates the two insulating
phases. We finally suppose that the $k_z$ value of the $P_{\text{C}2}$ plane
remains unchanged
by temperature, as it is unlikely that the Weyl nodes' annihilation point in
the BZ changes significantly from the static result.
We verified that the position of both
band extrema along the \mbox{A-L} line in the TI phase
does not change significantly over the range of
temperatures and pressures studied, between the
$P_\text{C1}$ and $P_\text{C2}$ planes. A more thorough
analysis of the
temperature dependence of the band-gap location within
the $P_{\text{C}1}$ and $P_{\text{C}2}$ planes, as well
as the pressure-temperature dependence of the Rashba
splitting, can be found in
Appendix~\ref{AppendixRashba}.

A more formal approach would also require one to redefine the topological
invariant for finite temperatures, relying on the density matrix rather than
on the static ground state wavefunction, since temperature modifies the
occupation of the electronic
eigenstates~\cite{budich_topology_2015,huang_topological_2014,
viyuela_two-dimensional_2014}. The question of the persistence of quantum
topological order at finite temperatures, compared to a nonquantized
\enquote{classical} topological order,  is still under
investigation~\cite{lu_singularity_2019}.

When considering the effect of temperature on the TPT within our framework,
there are thus two possible outcomes, which depend on the sign of the band-gap renormalization. In the first case, a negative correction implies that
the band gap closes with increasing temperature, which can both drive a
topologically trivial system towards a band inversion and stabilize an already
inverted band structure, thus globally promoting the topological phase. On the
other hand, a positive correction will favor the trivial phase: the band gap
opens with increasing temperature, preventing the band inversion in a trivial
system and destabilizing a TI band structure until the band inversion is
reversed. A more detailed version of this argument can be found in our
proceedings paper~\cite{brousseau-couture_qts_2019}.

In the case of a pressure-induced TPT with an intermediate WSM phase, both
critical pressures, $P_{\text{C1}}$ and $P_{\text{C2}}$, will be modified by
temperature: a negative renormalization favoring the topological phase will
diminish the amount of pressure required to close the band gap, while a
positive renormalization promoting the trivial phase will increase it. The
WSM phase width will also be affected by the relative strength of the
renormalization between the trivial and topological phases. The combined
effect of the strength and sign of the renormalization will determine if the
WSM phase is widened or narrowed by increasing temperature.

\subsubsection{Electron-phonon interaction}\label{EPIResults}

\begin{figure}
\centering
\includegraphics[width=\linewidth,trim={0.1cm 0 0.5cm 0},clip]{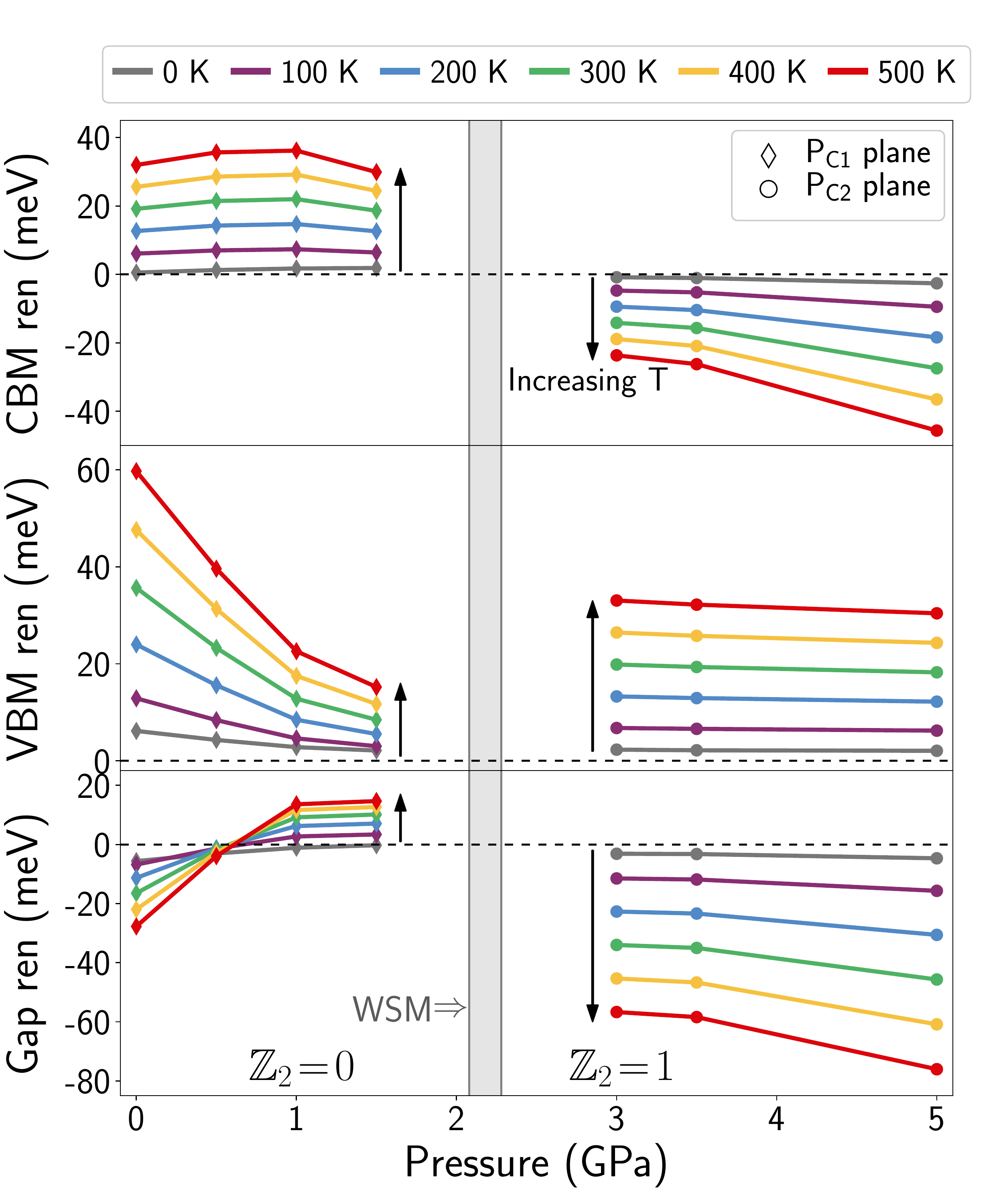}
\caption{\textbf{EPI contribution to the temperature-dependent
renormalization} for the conduction-band minimum (CBM, top panel), valence-band maximum (VBM, center panel) and total band gap (lower panel) of BiTeI,
as a function of pressure. A zero renormalization (black dashed lines) refers
to the static lattice results. The gray curves (closest to the dashed lines) show the zero-point
renormalization (ZPR). Black arrows indicate increasing temperature. For visual reference, the shaded gray area locates the
WSM phase boundaries from the static calculation.  }\label{FigBandRenormEPI}
\end{figure}

In a typical semiconductor, with a nearly parabolic dispersion near
well-defined band extrema and a sufficiently wide band gap, the Fan
contribution usually dominates the self-energy. One then expects from
Eq.~(\ref{EqnSigmaFan}) to observe a decrease of the band-gap energy with
increasing temperature: the VBM is repelled by neighboring occupied states
with similar but lower energy, leading to a positive renormalization.
Similarly, the CBM is repelled by neighboring unoccupied states with higher
energy, yielding a negative renormalization. Couplings between states with
different occupation factors are disadvantaged by the large energy difference
in the denominator, due to the presence of the gap. This behavior is sometimes
referred to in the literature as the Varshni
effect~\cite{cardona_isotope_2005}.

In BiTeI, the temperature dependence of the gap cannot be inferred by such a
heuristic argument, since, on the one hand, the small band gap emphasizes the
weight of couplings between the subsets of occupied and unoccupied bands
compared to a typical semiconductor, and, on the other, the Rashba splitting
creates regions in the BZ where a phonon with finite, nonvanishing
wavevector~$\mathbf{q}$ can couple electronic states with very similar
eigenenergies.

The EPI contribution to the VBM~(center panel), CBM~(top panel) and total
band gap~(bottom panel) is shown in Fig.~\ref{FigBandRenormEPI}.  In order to
track any temperature-induced change to the gap location and to accurately
capture the renormalization for the minimal gap, the EPI was computed in the
H-A-L plane for the trivial phase, and in the $P_{\text{C2}}$ plane for the
TI phase, for electronic states along the solid red and dashed blue paths shown in
Fig.~\ref{FigStructure}(b).

In the trivial insulator phase, both the VBM and the CBM are shifted towards
higher energies. While the CBM renormalization is almost pressure independent
for a given temperature, the VBM is strongly affected by pressure, its
renormalization dropping by roughly a factor of 3 between 0~and 1~GPa. This
behavior is most likely linked to the higher compressibility along the
$c$~axis at lower pressures~\cite{xi_signatures_2013}. The strength of this
variation could, however, be amplified by the overestimated compressibility
predicted by the PBE functional for BiTeI~\cite{guler-kilic_crystal_2015}. In
our case, this results in a total gap renormalization that changes sign within
the trivial phase, going from negative to positive shortly after 0.5~GPa.
Nevertheless, the EPI contribution to the gap remains quite small compared
to the bare gap energy at ambient pressure (at most 20~meV at all
temperatures). As argued before, such an opening of the gap with temperature
when approaching the TPT will delay the first gap closing, thus moving
$P_{\text{C1}}$ towards a slightly higher value.

In the TI phase, the renormalization displays a different behavior: the VBM
and CBM renormalizations have opposite signs, both contributing to a decrease
of the gap energy with increasing temperature. Again, this behavior is not
favorable for the nontrivial case as it works to revert the already inverted
gap, driving the system back to the trivial phase. One should also note that,
in this phase, it is the VBM that exhibits a pressure-independent behavior,
while, for the CBM, the negative renormalization increases steadily with
increasing pressure. This seemingly distinct behavior is simply a signature
of the band inversion: it can be understood by recalling that the orbital
decomposition of the state being corrected at the VBM in the TI phase has the
same character as the CBM in the trivial phase, as was emphasized in
Fig.~\ref{FigFatbands}. Similarly, the CBM in the trivial phase has the same
leading orbital character as the VBM in the TI phase. From this point of view,
the band extremum dominated by Bi-$p_z$ character exhibits a quasi-pressure-independent renormalization, while the extremum with Te/I-$p_z$
character is gradually shifted towards lower energies throughout the studied
pressure range. It is also rather peculiar that only the Te/I states are
affected by the change of topology, which can be seen in the
temperature-dependent renormalization at a given pressure going from positive
to negative as the system goes through the phase transition. A more thorough
explanation of this sign change would require one to probe the EPI in the
gap-closing region,  
which was not possible in the scope of this paper. Nevertheless, if considering
the gap behavior in both phases, we can infer that the EPI globally promotes
the trivial phase, pushing the TPT towards higher pressures as temperature
increases.



\subsubsection{Thermal expansion}\label{TEResults}

\begin{figure}
\centering
\includegraphics[width=\linewidth, trim={0.2cm 0 0.5cm 0},clip]{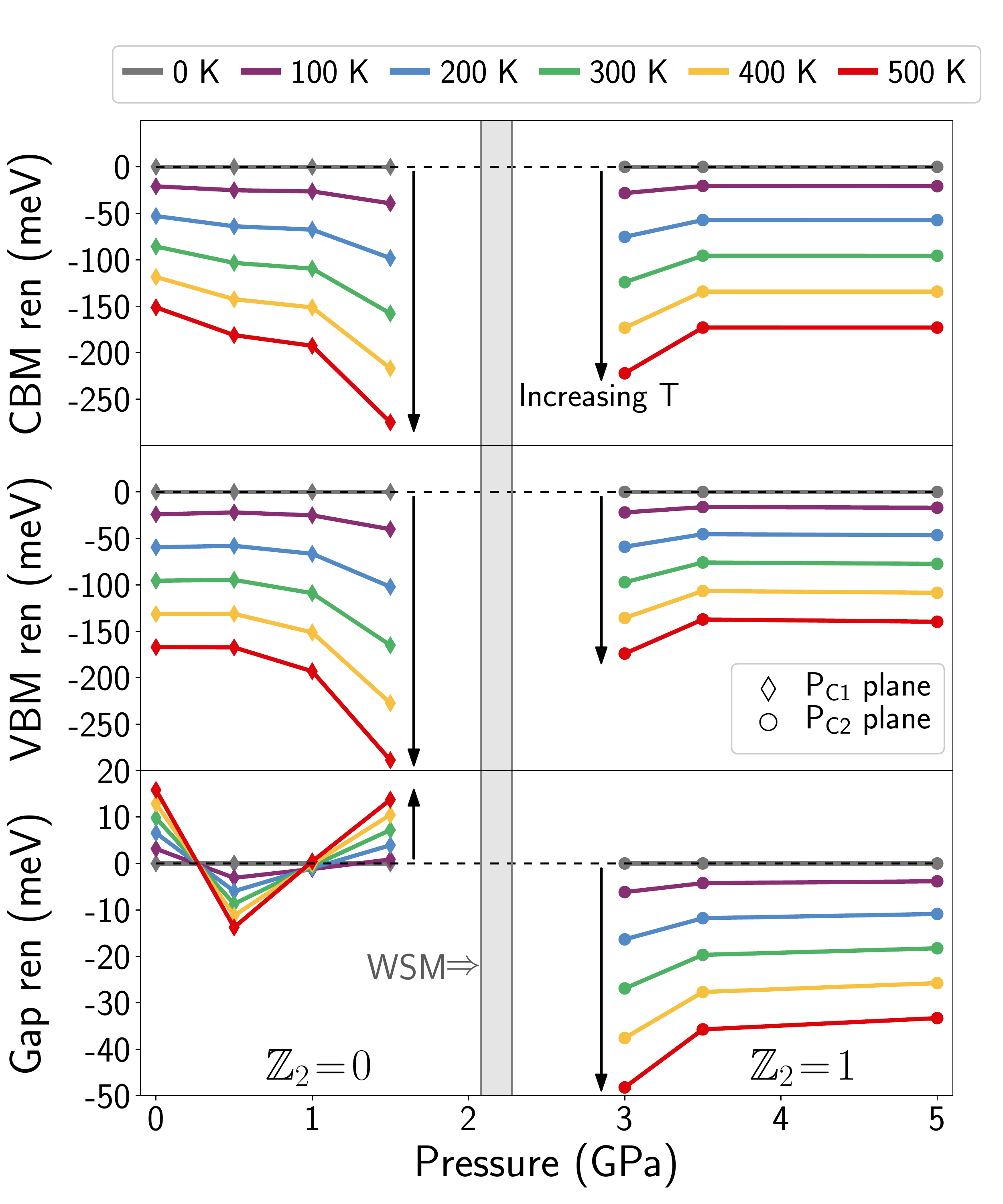}
\caption{\textbf{TE contribution to the temperature-dependent renormalization}
for the conduction-band minimum (CBM, top panel), valence-band maximum (VBM,
center panel) and total band gap (lower panel) of BiTeI, as a function of
pressure. A zero renormalization (black dashed lines) refers
to the static lattice results. 
Black arrows indicate increasing
temperature. For
visual reference, the shaded gray area locates the WSM phase boundaries from
the static calculation. Note that the Gr\"uneisen formalism neglects the
zero-point phonon vibrational energy, hence the identically zero correction
at $0$~K (gray).} \label{FigBandRenormTE}
\end{figure}

The TE contribution to the total renormalization is
displayed in Fig.~\ref{FigBandRenormTE}, for the CBM (top panel), VBM (center
panel), and indirect band gap (bottom panel). When solely looking at the band
renormalization, one finds a seemingly monotonic behavior, both bands being
shifted towards lower absolute energies. The pressure-dependent
renormalization at a given temperature does not seem affected by the leading
orbital character of the band extrema, as was the case for EPI. In the TI
phase, the renormalization rate at a given pressure is greater for the CBM
than for the VBM, hence a global decrease of the gap energy with increasing
temperature. In the trivial phase, the indirect band-gap renormalization shows
a far more erratic behavior: while for most pressures the renormalization rate
is greater for the VBM than for the CBM, hence a gap opening with increasing
temperature, at 0.5~GPa we obtain the opposite trend. This is reminiscent of
the sign change observed in this pressure range for the EPI contribution [see
Fig.~\ref{FigBandRenormEPI}, lower panel], which we again attribute to an
overestimated compressibility in the low-pressure regime.
For a layered structure like BiTeI, a higher compressibility will mainly
enhance the $s_{33}$ compliance constant entering the definition of $\alpha_c$
(Eq.~(\ref{EqnAlpha})), which will, in turn, modify the thermal expansion
along this axis and the resulting gap renormalization.
As discussed more thoroughly in %
Appendix~\ref{AppStructural},
the lattice structure is reasonably
well described by PBE for $P\geq 1$~GPa. Thus, in the following section, we
only consider the normalization trends for pressures of 1~GPa and higher to
evaluate the temperature-dependent critical pressures. One should finally note
that the TE contribution has the same order of magnitude as the EPI
contribution, such that it cannot be neglected when investigating the
temperature dependence of the gap, as it is often done for materials with
light ions.

We recall here that, as mentioned in Section~\ref{theoryTE}, the Gr\"uneisen
formalism neglects the zero-point phonon vibrational energy, such that the
${T=0}$~K lattice parameters are identical to the static ones. We estimated the
missing zero-point thermal expansion contribution to the band-gap
renormalization using the high-temperature extrapolation method displayed in
Fig.~2 of Cardona and Thewalt~\cite{cardona_isotope_2005}. In this method, the
renormalized lattice parameters
at ${T=0}$~K are evaluated by extrapolating the high-temperature slope of
$\Delta a(T)/a$ and $\Delta c(T)/c$ to $0$~K. This led to a nearly constant
contribution of $\sim2$~meV in the trivial phase and $\sim -5$~meV in the
topological phase. This missing contribution does not affect the qualitative
trends previously described.

The renormalization trends observed when approaching the TPT in both
insulating phases, a gap opening in the trivial phase and a gap closing in the
topological phase, are precisely the trends observed for the EPI contribution.
Consequently, just like EPI, TE delays the TPT, moving both critical pressures
towards higher values with increasing temperature. Excluding $P<1$~GPa from
the analysis for previously stated reasons, these trends agree qualitatively
with Monserrat and Vanderbilt~\cite{monserrat_temperature_2017}, despite being
smaller in absolute value. This could partly be attributed to the fact that
Monserrat and Vanderbilt minimized the Gibbs free energy independently for
each lattice parameter, while our methodology [see Eq.~(\ref{EqnGammaHex}],
correlates the expansion along a given axis with the \enquote{thermal
pressure}~\cite{munn_role_1972} induced on the lattice by varying both lattice
parameters, through $\gamma^{a1}$ and $\gamma^c$.


This discrepancy could also be explained by anharmonic contributions to the
phonon free energy, which our perturbative methodology does not capture. A
more refined and numerically accurate calculation of the anisotropic thermal
expansion coefficients would rely on the full temperature-dependent
phonon-perturbed potential, taking into account all anharmonic interactions,
as implemented in the Temperature Dependent Effective Potential method (TDEP) method~\cite{pike_calculation_2019}.


\subsection{Weyl semimetal phase evolution}\label{WSMResults}

\begin{figure}
\centering
\includegraphics[width=\linewidth, trim={0.6cm 5.5cm 1cm 0},clip]{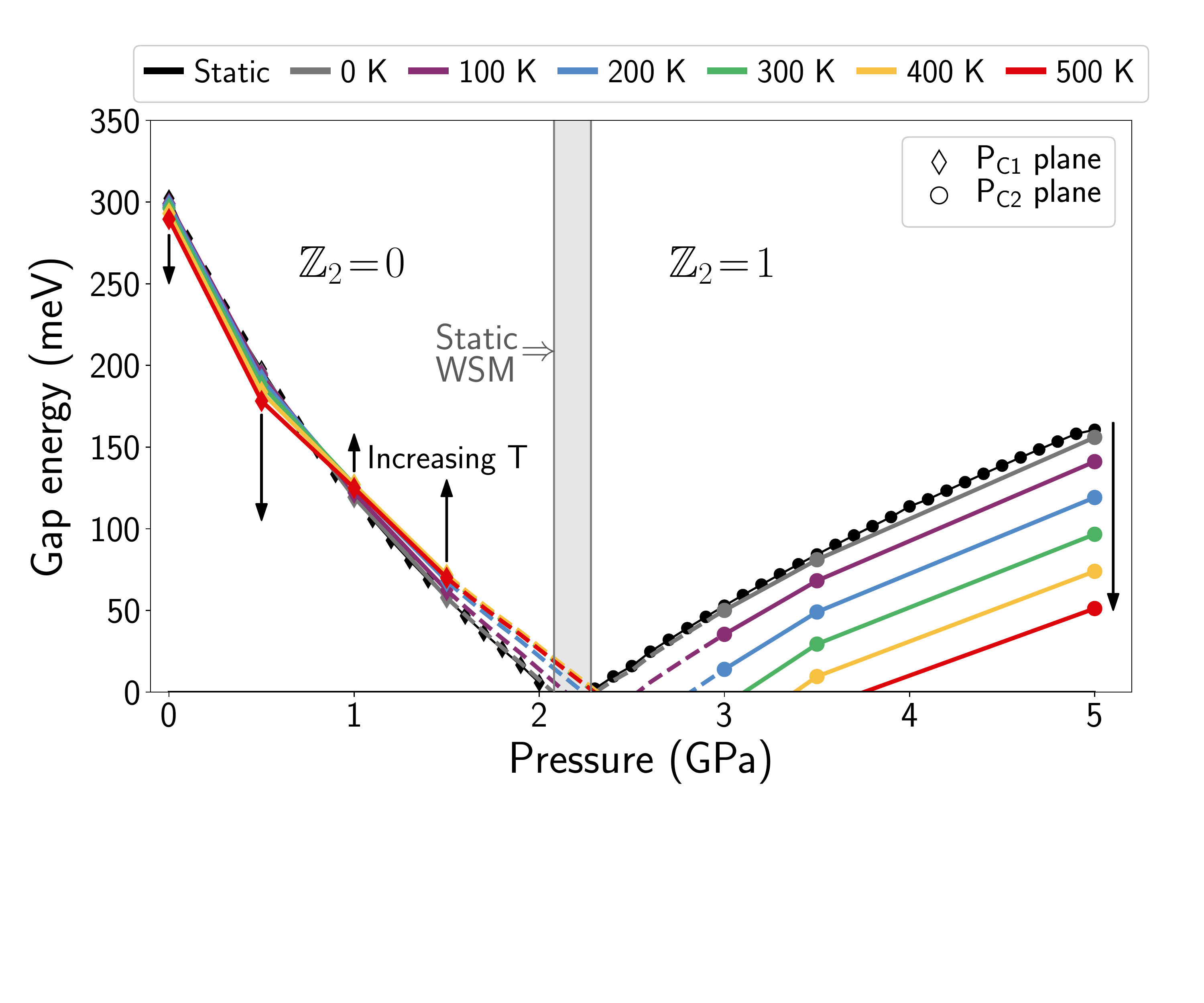}\\
\caption{\textbf{Temperature-dependent band-gap energy.} At a given
temperature, the total renormalization of the band gap (EPI+TE, solid-colored
lines) is extrapolated towards the TPT region and applied to the bare gap
energies (black markers with denser pressure sampling). The resulting temperature-dependent gap energies
within the pressure region excluded from our calculations are shown with
colored dashed lines. Black arrows indicate increasing temperature. For visual reference, the static lattice WSM phase is
shown in shaded gray. }\label{FigGapPT}
\end{figure}

\begin{figure}
\centering
\includegraphics[width=\linewidth,trim={0.6cm 2.5cm 1.2cm
0},clip]{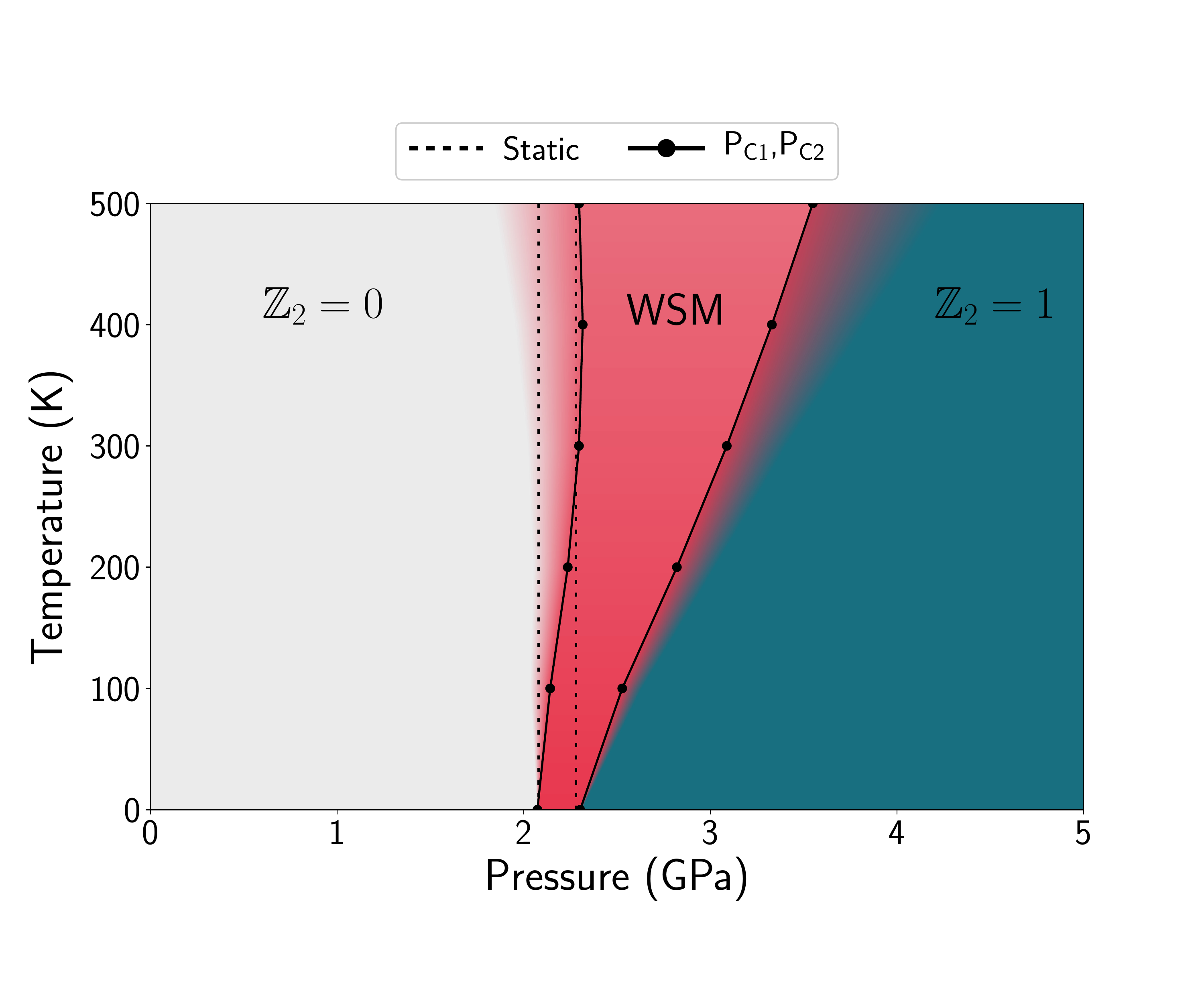}
\caption{\textbf{Phase diagram of BiTeI.} The temperature-dependent WSM phase
boundaries are the intersections of the extrapolated gap closing energies from
Fig.~\ref{FigGapPT}. Both critical pressures are pushed towards higher values.
The renormalization is stronger for $P_{\text{C2}}$ than for $P_{\text{C1}}$,
resulting in a widening of the WSM phase with increasing temperature. At each
temperature, we defined a crossover region between topologically distinct
phases, since one cannot physically distinguish an insulating phase from a
gapless phase if $E_{g}<k_B T$, due to thermal excitations. See
Section~\ref{WSMResults} for more details.} \label{FigPTDiagram}
\end{figure}

The temperature dependence of the WSM phase width was assessed by combining
the results displayed in the two previous sections to evaluate the full gap
energy renormalization at each pressure and temperature. For each
$\mathbf{k}$~point on the paths considered [see Fig.~\ref{FigStructure}(b)],
the EPI and TE corrections were added to the static band structure before
computing the resulting temperature-dependent band gap.

Since, for technical reasons discussed in Section~\ref{staticresults}, we
could not directly probe the EPI in the gap closing region, we tracked the
temperature dependence of $P_{\text{C1}}$ and $P_{\text{C2}}$ by extrapolating
the total renormalization behavior for each band extrema from neighboring
pressures towards the TPT in the trivial and topological phases. This
correction was then added to the bare gap energy computed with the static
lattice (black markers with denser pressure sampling), as shown by the colored dashed lines in
Fig~\ref{FigGapPT}.


The temperature-dependent critical pressures extracted from
Fig.~\ref{FigGapPT} are explicitly shown in the pressure-temperature phase
diagram of Fig.~\ref{FigPTDiagram} (black markers). For visual reference, the critical
pressures obtained with the static lattice are shown with dotted lines. The
numerical values for $P_{\text{C}1}(T)$ and $P_{\text{C}2}(T)$ are detailed
in Table~\ref{TableWSM} of Appendix~\ref{AppendixWSM},
as well as the WSM phase boundaries obtained
by considering only the contribution from EPI or TE.

As discussed in the previous sections, both EPI and TE are unfavorable to
the topological phase. We therefore observe without surprise that both
$P_{\text{C1}}$ and $P_{\text{C2}}$ are shifted towards higher pressures.
However, as the renormalization from both contributions is stronger in
magnitude in the topological phase, the temperature-dependent variation is
much weaker for $P_{\text{C1}}$ than for $P_{\text{C2}}$. Thus, the WSM phase
width increases with temperature. This effect is already sizable at ${T=0}$~K,
where the zero-point motion of the ions induces a 15\% increase in the WSM
phase width, compared to the static lattice. It has almost doubled by
$T=100$~K, despite the absolute WSM phase width remaining small at 0.39~GPa.

The only deviation to this behavior is the slight decrease of $P_{\text{C}1}$
above room temperature. This can be explained by the fact that, for $P=1$~and
1.5~GPa, EPI and TE do not shift the band extrema in the same way in the BZ.
Therefore, in this pressure range, the total renormalization does not sum up
to the individual corrections.


One should also take into consideration the fact that, at finite temperature,
thermal excitations do not allow one to physically distinguish between an
insulating and a gapless phase, if the band-gap energy is smaller than
$k_B T$. Thus, for each temperature, we defined a crossover region between
topologically distinct phases by evaluating the pressure difference between
a given static lattice critical pressure and the pressure at which
$E_g = k_\text{B}T$ in the appropriate insulating phase, using the data from
Fig.~\ref{FigBareGapP}. In Fig.~\ref{FigPTDiagram}, these regions identify
the phase space where, within our methodology, the topological index cannot
be properly assessed. Lastly, we also added to Fig.~\ref{FigPTDiagram} a color
intensity gradient in the high-temperature regime of the WSM phase, to
emphasize the fact that our calculations did not explicitly treat the
temperature dependence within this pressure region. Therefore, there remains
some uncertainty about the topological nature of BiTeI in this region of the
pressure-temperature phase space.


From these results, we can conclude that temperature effects do not restrain
the experimental detection of the WSM phase. In the low-temperature regime,
where quantum effects can more easily be observed thanks to the reduction of
thermal noise, the WSM phase widening remains however small, despite being
sizable when compared to the static lattice phase width. The experimental
signatures of the TI phase will be found at higher pressures with increasing
temperature. These qualitative temperature dependencies can assist the
experimental effort when designing experiments the purpose of which is to detect
signatures of a topologically nontrivial phase. Moreover, they shed some
light on the subtle interplay between EPI and TE and their effect on the phase
space of topological phases in noncentrosymmetric materials. Since the
presence of gapless modes characterizes both the WSM and $\mathbb{Z}_2$ TI
phases, additional experiments focusing on observables that can physically
distinguish between these two nontrivial phases, and that are feasible under
isotropic hydrostatic pressure, could provide more insight about this
phenomenon.\\

\subsection{Discussion}\label{discussion}

The purpose of this paper was to determine the qualitative trend of the
temperature-dependent renormalization of $P_{\text{C1}}$ and $P_{\text{C2}}$: the global
behavior of the TPT, more than the precise numerical values of the critical
pressures, was our main target. From this point of view, there
are several limitations to our analysis, the first of them being the
$\mathbf{q}$-point sampling for the EPI calculation. BiTeI is a polar
material, i.e., has nonvanishing Born effective charges, with
infrared-active phonon modes. At long wavelength, ${|q|\rightarrow0}$, atomic
displacements along longitudinal optical (LO) modes generate a macroscopic
electric field throughout the material, which can couple to the electrons
through dielectric interaction. This particular type of EPI, known as the
Fr\"ohlich interaction~\cite{frohlich_electrons_1954}, has been shown to
dominate the ZPR for polar
materials~\cite{nery_influence_2016,miglio_generalized_2019} and to cause an
unphysical divergence of the ZPR within the adiabatic AHC
framework~\cite{ponce_temperature_2014}, in which Eq.~(\ref{EqnSigmaFan}) is
approximated by neglecting the phonon frequencies in the denominators.

Thus, for polar materials, evaluating the precise contribution of the
long-range Fr\"ohlich interaction to the ZPR within the nonadiabatic AHC
framework requires a very dense $\mathbf{q}$-point grid, which for technical
reasons was not available to us at the time of this study.
The methodology developed by Nery and Allen~\cite{nery_influence_2016} to
estimate the polar contribution missing from a coarse $\mathbf{q}$-point grid
calculation could also not be straightforwardly applied to BiTeI for two
reasons. First, the
band extrema are not located at the zone center and spin-orbit coupling is
mandatory to describe Rashba splitting correctly. Second, the validity of the
parabolic effective mass approximation near the band extrema is also
questionable for this material, especially at higher pressures.
Our results should consequently be understood as a lower bound to the full
EPI contribution.


The extrapolation technique used to evaluate the temperature dependence of
the critical pressures is also by itself limited by the assumption that the
renormalization behavior we observe in both insulating phases varies
monotonically towards the TPT. This assumption could only be validated by an
explicit calculation of the EPI in the close vicinity of the TPT, which would
require one to lift the diagonal self-energy approximation, as excitations across
the gap will no longer be negligible. The addition of these nondiagonal
coupling terms to the self-energy, which we conjecture will have a significant
contribution, could alter the general trends observed in Fig.~\ref{FigGapPT}.
These extra terms will also be crucial to fully characterize the effect of
EPI on the Te/I-$p_z$ band extrema near the critical pressures (VBM for
$\mathbb{Z}_2=0$, CBM for $\mathbb{Z}_2=1$; see Fig.~\ref{FigBandRenormEPI}
and Section~\ref{EPIResults}).

Lastly, our analysis supposed a perfect semiconductor crystal, while
experimentally BiTeI has been shown to be naturally $n$-doped due to small
deviations in stoichiometry~\cite{makhnev_optical_2014}. The presence of such
doping charge could slightly affect the phonon-perturbed crystal potential,
which could, in return, modify the matrix elements entering the EPI
self-energy. We do not however expect this self-doping to have a strong impact
on the qualitative trends observed throughout this paper, as long as the amount
of defects remains small compared to substitutional doping.


\section{Conclusion}\label{conclusion}
In the present paper, we have characterized the temperature dependence of the
topological phase transition in BiTeI using first-principles methodologies
based on DFPT. The electron-phonon interaction was obtained with the
nonadiabatic AHC methodology, while the thermal expansion was computed using
the Gr\"uneisen parameters formalism.

The indirect band-gap renormalization induced by EPI changes sign as the
system undergoes the phase transition, opening the gap in the trivial phase
and closing it in the TI phase. The band extremum with leading Bi-6$p_z$
character displays a quasi pressure-independent behavior, while the
Te/I-$5p_z$ extremum manifests a strong sensitivity to both pressure and
topology.

The thermal expansion contribution to the band-gap renormalization has the
same order of magnitude as the EPI contribution. At a given temperature, both
band extrema are affected in a similar manner throughout the pressure range,
regardless of their leading orbital character. The resulting band-gap
renormalization nevertheless captures the change of topology: it mainly
increases the indirect gap in the trivial phase and reduces it throughout
the topological phase.

The combined effect of EPI and TE globally moves the TPT of BiTeI towards
higher pressures. The resulting temperature-dependent renormalization is
stronger in magnitude in the TI phase compared to the trivial phase, such that
the intermediate WSM phase is widened by temperature. Clear signatures of the
$\mathbb{Z}_2$ topological insulator phase should, therefore, appear at
higher pressures as temperature is increased.

Overall, our results indicate that temperature effects are not negligible for
materials with heavy ions and must be accounted for when engineering devices
relying on topological properties. Our findings can also aid the search for
experimental evidence of the topologically nontrivial phases of BiTeI, and
more generally reveal how temperature can have a substantial influence on the
phase space of topological phases of noncentrosymmetric materials.\\
\par \emph{\textbf{Note added.}} Recently, we learned of an article by Lihm and
Park\cite{lihm_phonon-induced_2020} which is relevant
to this research field.

\begin{acknowledgments}

This research was financially supported by the Natural Sciences and
Engineering Research Council of Canada (NSERC), under the Discovery Grants
program grant No. RGPIN-2016-06666. Computations were made on the
supercomputers Beluga and Briaree managed by Calcul Qu\'ebec and Compute Canada.
The operation of these supercomputers is funded by the Canada Foundation for Innovation (CFI), the Ministère de l’Économie, de la Science et de l’Innovation du Québec (MESI) and the Fonds de recherche du Québec - Nature et technologies (FRQ-NT). V.B.-C. acknowledges support by the NSERC Alexander
Graham Bell Canada Graduate Scholarship doctoral program.

\end{acknowledgments}

\appendix 
\section{Effect of Van der Waals dispersion correction
}\label{AppendixStructure}

Since the first prediction of the TPT in
BiTeI~\cite{bahramy_emergence_2012},
numerous studies of ground-state properties of bismuth tellurohalides
(BiTeX,
X=Cl, Br, I) have appeared in the first-principles
literature~\cite{rusinov_many-body_2013,rusinov_role_2015,
rusinov_pressure_2016,sklyadneva_lattice_2012,eremeev_two-_2017,
schwalbe_ab_2016}. Many of them have discussed the fact that PBE-GGA
does not
capture the Van der Walls bonding properly along the normal axis for this
class of materials, resulting in a general overestimation of the lattice
parameters~\cite{rusinov_pressure_2016,guler-kilic_crystal_2015},
up to 7\%
for the $\hat{c}$ axis~\cite{rusinov_pressure_2016}. PBE-GGA was also
criticized as yielding an unrealistic compressibility for BiTeI at low
pressures~\cite{guler-kilic_crystal_2015}. It has also been shown that
dispersion-corrected DFT, which accounts for Van der Waals forces
through a
semi-empirical dispersion potential, reproduces more accurately the
pressure
dependence of the lattice structure for bismuth
tellurohalides~\cite{guler-kilic_pressure_2016}. These studies did not,
however, investigate the impact of these corrections on the phonon
frequencies,
nor on the electron-phonon interaction. In the following, we include
the Van
der Waals interaction within the  DFT framework, using
dispersion-corrected
DFT-D3BJ, where a semi-empirical dispersion potential is
added~\cite{grimme_consistent_2010} in conjunction with Becke-Jonhson
damping~\cite{johnson_post-hartree-fock_2006,grimme_effect_2011} to avoid
short-range diverging behavior. DFT-D3BJ was shown to be one of the most
reliable dispersion-corrected methods to describe vibrational
properties~\cite{van_troeye_interatomic_2016}. Both the trivial (0,
1.5~GPa)
and the topological (3.5, 5~GPa) phases are considered.

\subsection{Structural properties}\label{AppStructural}

\begin{table*}
\setlength\extrarowheight{4pt}
\caption{\textbf{Static lattice properties at
0~and~5~GPa.}}\label{TableLattice}
\begin{tabular}{|c||c|c|c||c|c|c|}
\hline\hline
&\multicolumn{3}{c||}{0~GPa}&\multicolumn{3}{c|}{5~GPa}\\
\cline{2-7}
& Exp. & PBE & DFT-D3BJ & Exp. & PBE & DFT-D3BJ \\[2pt]
\hline
a (\AA) & 4.34~\cite{shevelkov_crystal_1995} & 4.44 & 4.36 
& 4.15~\cite{xi_signatures_2013} & 4.23 & 4.12 \\

c (\AA) &  6.85~\cite{shevelkov_crystal_1995} & 7.31 & 6.78 &
6.52~\cite{xi_signatures_2013} & 6.55 & 6.49 \\

c/a ratio & 1.58~\cite{shevelkov_crystal_1995} & 1.65 & 1.56 &
1.57~\cite{xi_signatures_2013} & 1.58 & 1.58 \\

gap (eV) & 0.38~\cite{ishikaza_giant_2011}, 0.36~\cite{lee_optical_2011},
0.33~\cite{makhnev_optical_2014} & 0.30 & 0.03
&  & 0.16 & 0.30 \\
Bi-Te $\hat{c}$ (\AA) & 2.10~\cite{shevelkov_crystal_1995} & 1.70 &
1.73 & 
& 1.75 & 1.78 \\

Bi-I $\hat{c}$ (\AA) & 
1.72~\cite{shevelkov_crystal_1995} & 2.11 & 2.15 & & 2.16 & 2.17\\
\hline \hline
\end{tabular}
\end{table*}
\begin{table*}
\setlength\extrarowheight{4pt}
\caption{\textbf{Optimized static lattice structure}}\label{TableOpt}
\begin{tabular}{|c||c|c|c|c|c|c|c|}
\hline\hline
& 0~GPa & 0.5~GPa & 1~GPa & 1.5~GPa & 3~GPa & 3.5~GPa & 5~GPa \\[2pt]
\hline
a (\AA) & 4.44 &4.41 &4.39 &4.37 &4.30 &4.28 & 4.23\\

c (\AA) &  7.31 &7.05 &6.90 &6.81 &6.66 &6.62 & 6.55\\
Bi-Te $\hat{c}$ (\AA) &  1.70 & 1.71&1.71 &1.72 &1.74 &1.75 & 1.75\\

Bi-I $\hat{c}$ (\AA) & 
2.11  &2.12 &2.13 &2.14 &2.15 &2.15 & 2.16 \\
\hline \hline
\end{tabular}
\end{table*}

\begin{table}
\setlength\extrarowheight{4pt}
\caption{\textbf{Static lattice 0~GPa Rashba parameters along A-L}. }\label{TableRashba}
\begin{tabular}{|c|c|c|c|}
\hline\hline
 & Exp. & PBE &DFT-D3BJ\\[2pt]
\hline
 E$_\text{R}$ (meV) & 100~\cite{ishikaza_giant_2011}
 108~\cite{sakano_strongly_2013}& 103 & 181\\
 $k_{\text{R}}$ (\AA$^{-1}$) & 0.052~\cite{ishikaza_giant_2011},
 0.046~\cite{martin_quantum_2013}, 0.050~\cite{sakano_strongly_2013}&
 0.050&0.053\\
 $\alpha_{\text{R}}$ (eV$\cdot$\AA) & 3.85~\cite{ishikaza_giant_2011}
 4.3~\cite{sakano_strongly_2013}& 4.13 & 6.86\\ 
\hline \hline
\end{tabular}
\end{table}

The lattice parameters, bare band gap energies and equilibrium 
distance between
the Te/I and Bi planes for 0~and~5~GPa are shown in
Tables~\ref{TableLattice}
and~\ref{TableRashba}, with comparison to available experimental data.
In good
agreement with previous
works~\cite{monserrat_temperature_2017,rusinov_pressure_2016,
guler-kilic_crystal_2015}, PBE-GGA overestimates the lattice parameters at
ambient pressure more than what is typically expected from DFT
calculations,
at 2.3\% for $a$ and 6.7\% for $c$, while the dispersion-corrected method
reproduces  more accurately the experimental structure. At 5~GPa, both
methods
agree within 2\% of experimental value, for both $a$ and $c$. The
interplanar
equilibrium distances at both pressures are in good agreement with
experiment.
Note that the positions of Te and I planes are inverted when compared 
to the
experimental results~\cite{shevelkov_crystal_1995}; this effect has been
attributed to the fact that x-ray diffraction measurements cannot
establish a
clear distinction between the Te and I planes, due to their very
similar atomic
radii and charge~\cite{bahramy_origin_2011}. This inversion is systematic
throughout the first-principles
literature~\cite{monserrat_temperature_2017,guler-kilic_crystal_2015,
sklyadneva_lattice_2012,bahramy_origin_2011}.

Analyzing the full pressure dependence of the lattice parameters for
PBE-GGA
and DFT-D3BJ between 0~and 5~GPa (not included here) lead to similar
results
as those of G\"uler-Kili\c{c} and
Kili\c{c} (see Figs.~3(g) and 3(h) of~\cite{guler-kilic_crystal_2015}), where the overestimation of $a$ remains around 2\%
throughout
the whole pressure range, while for $c$ it drops rapidly under 2\%
after 1~GPa.
Thus, for $P\gtrsim 1$~GPa, the lattice structure is similarly well
described by both methods. The resulting lattice parameters and interatomic distances along $\hat{c}$ obtained by the optimization procedure are presented in Table~\ref{TableOpt} for all pressures included in this work.

The bulk band gap of BiTeI has been measured at ambient pressure via
ARPES~\cite{ishikaza_giant_2011}, optical
spectroscopy~\cite{lee_optical_2011}
and optical spectral ellipsometry~\cite{makhnev_optical_2014}, yielding
respectively 0.38, 0.36~and 0.33~eV. Another soft x-ray ARPES experiment
obtained a smaller value of 0.26~eV~\cite{sakano_three-dimensional_2012};
they however argued that their measurement probes a region between
the bulk
and the subsurface, where some band bending effects could lower
the measured
band gap compared to the bulk value. Our results with PBE, 0.30~eV for the
indirect gap and 0.33~eV for the direct gap, are in pleasantly good
agreement
with experiment, thanks to the overestimation of the lattice parameters
counterbalancing PBE's well-know gap
underestimation~\cite{rusinov_pressure_2016}. For DFT-D3BJ, the band 
gap is strongly underestimated: the system is almost metallic, with
a band gap of
about 30~meV at ambient pressure. We are not aware of any reference for
experimental gap measurements in the topological phase.

Beyond the crystal structure, the parameters characterizing the Rashba
splitting can also be used to assess the level of agreement of the
calculated electronic structure with experiment. The Rashba energy,
$\text{E}_\text{R}$, is defined as the energy difference between
electronic
states at the band extrema and at the time-reversal protected
degeneracy point
of spin-split bands, namely A for BiTeI [see
Fig.~\ref{FigFatbands}].
The momentum offset between those two $\mathbf{k}$-points is
captured
by the Rashba wavevector, $\mathbf{k}_\text{R}$. The Rashba parameter,
\begin{equation}\label{EqnRashba}
    \alpha_\text{R} = \frac{2 E_\text{R}}{k_\text{R}},
\end{equation}
is the strength of the
spin-orbit interaction in the Rashba Hamiltonian. These parameters are
clearly
displayed in the left panel of Fig.~2a and Eq.~(1) of Monserrat and
Vanderbilt~\cite{monserrat_temperature_2017}. Our calculations agree
reasonably
well with available experimental data [see Table~\ref{TableRashba}],
with a clear advantage for PBE-GGA. 

\subsection{Vibrational properties}

\begin{table*}
\setlength\extrarowheight{3pt}
\caption{\textbf{Zone-center optical phonon frequencies $\omega$
(cm$^{-1}$)
in the trivial phase}. When available, LO-TO splitting is indicated with
parentheses. LO-TO splitting occurs when the applied electric field is
in plane
for E modes, and normal for A$_1$ modes. }\label{TablePhononfreqZ0}
\begin{tabular}{|c||c|c|c|c||c|c|c|}
\hline\hline
&\multicolumn{4}{c||}{0~GPa}&\multicolumn{3}{c|}{1.5~GPa}\\
\cline{2-8}
Mode & Exp.\cite{sklyadneva_lattice_2012};\cite{gnezdilov_enhanced_2014};
\cite{ponosov_dynamics_2014};\cite{tran_infrared-_2014} &
Calc.\cite{rusinov_role_2015};\cite{sklyadneva_lattice_2012} & PBE  & 
DFT-D3BJ& Exp.\cite{ponosov_dynamics_2014};\cite{tran_infrared-_2014}
& PBE 
& DFT-D3BJ\\[3pt]
\hline
E$^1$  & 52.5; 55; 56; n/a  & 59; 54 (73) & 53 (74) & 57 (77) & 60; n/a & 
58 (77) &  62 (80) \\
A$_1^1$    & 90.5; 93; 92; 92 & 92; 88 (94) & 90 (94) &  90 (100)  &
96; 96 & 92 (100) & 95 (106)\\
E$^2$    & 99; 102; 101; 101& 101; 96 (115) &97 (118) &  101 (117)  &
106; 107 & 102 (119) & 106 (118)  \\
A$_1^2$    & 146.5; 150; 148; 146 & 152; 141 (142) & 141 (142) &
146 (148) &
152; 153 & 147 (148) &   151 (152) \\
\hline\hline
\end{tabular}
\end{table*}

\begin{table*}
\setlength\extrarowheight{3pt}
\caption{\textbf{Zone-center optical phonon frequencies $\omega$
(cm$^{-1}$)
in the topological phase}. When available, LO-TO splitting is indicated
with
parentheses. LO-TO splitting occurs when the applied electric field is
in plane
for E modes, and  normal for A$_1$ modes.  }\label{TablePhononfreqZ1}
\begin{tabular}{|c||c|c|c||c|c|c|}
\hline\hline
&\multicolumn{3}{c||}{3.5~GPa}&\multicolumn{3}{c|}{5~GPa}\\
\cline{2-7}
Mode & Exp.\cite{ponosov_dynamics_2014};\cite{tran_infrared-_2014} &
PBE  &
DFT-D3BJ& Exp. \cite{ponosov_dynamics_2014};\cite{tran_infrared-_2014}& 
PBE & DFT-D3BJ\\[3pt]
\hline
E$^1$  & 66; n/a & 63 (81) & 67 (84)  & 70$^{a}$; n/a & 68 (84) &   71 (87) \\
A$_1^1$    & 105; 104 & 99 (108) &102 (111)  & 107; 108 & 102 (114) &
109 (121)\\
E$^2$    & 112; 112 &108 (120) & 115 (120)  & 115; 115 & 111 (122) & 
115 (123)\\
A$_1^2$    & 162; 162 & 154 (155) & 159 (151)  & 165; 167 & 159 (160) &
164 (166)\\
\hline\hline
\multicolumn{5}{l}{$^{a}$\footnotesize{Extrapolated to 5~GPa from the
referenced experimental data.}}
\end{tabular}
\end{table*}

At the zone center, $\Gamma$, BiTeI has $C_{3v}$ point group symmetry,
yielding four Raman-active modes, whose irreducible representations
are two
doubly-degenerate $E$ modes and two $A_1$
modes~\cite{gnezdilov_enhanced_2014,tran_infrared-_2014}. Since BiTeI
has no
inversion center, all modes are both Raman and infrared
active~\cite{tran_infrared-_2014}.
Phonon frequencies for these modes are compared to available
experimental data
and previous first-principles calculations in
Tables~\ref{TablePhononfreqZ0}
and~\ref{TablePhononfreqZ1}. At ambient pressure, we obtain a relative
mean
deviation with experiment of 4.1\%  for PBE and 1.6\% for D3BJ. At 5~GPa,
these deviations drop to 3.8\% for PBE and 0.8\% for D3BJ. We verified
that
the general pressure dependence of the Raman-active frequencies agreed
well
with experiment in both cases. Our calculated values for LO-TO splitting
are in excellent agreement with previous
calculations~\cite{sklyadneva_lattice_2012}. A small feature measured at
$\omega$=118.5 cm$^{-1}$ at ambient
pressure~\cite{sklyadneva_lattice_2012}
can be associated with the LO-TO split E$^2$ mode, thus validating our
results.

\subsection{Discussion}
\begin{figure}
    \centering
    \includegraphics[width=\linewidth]
    {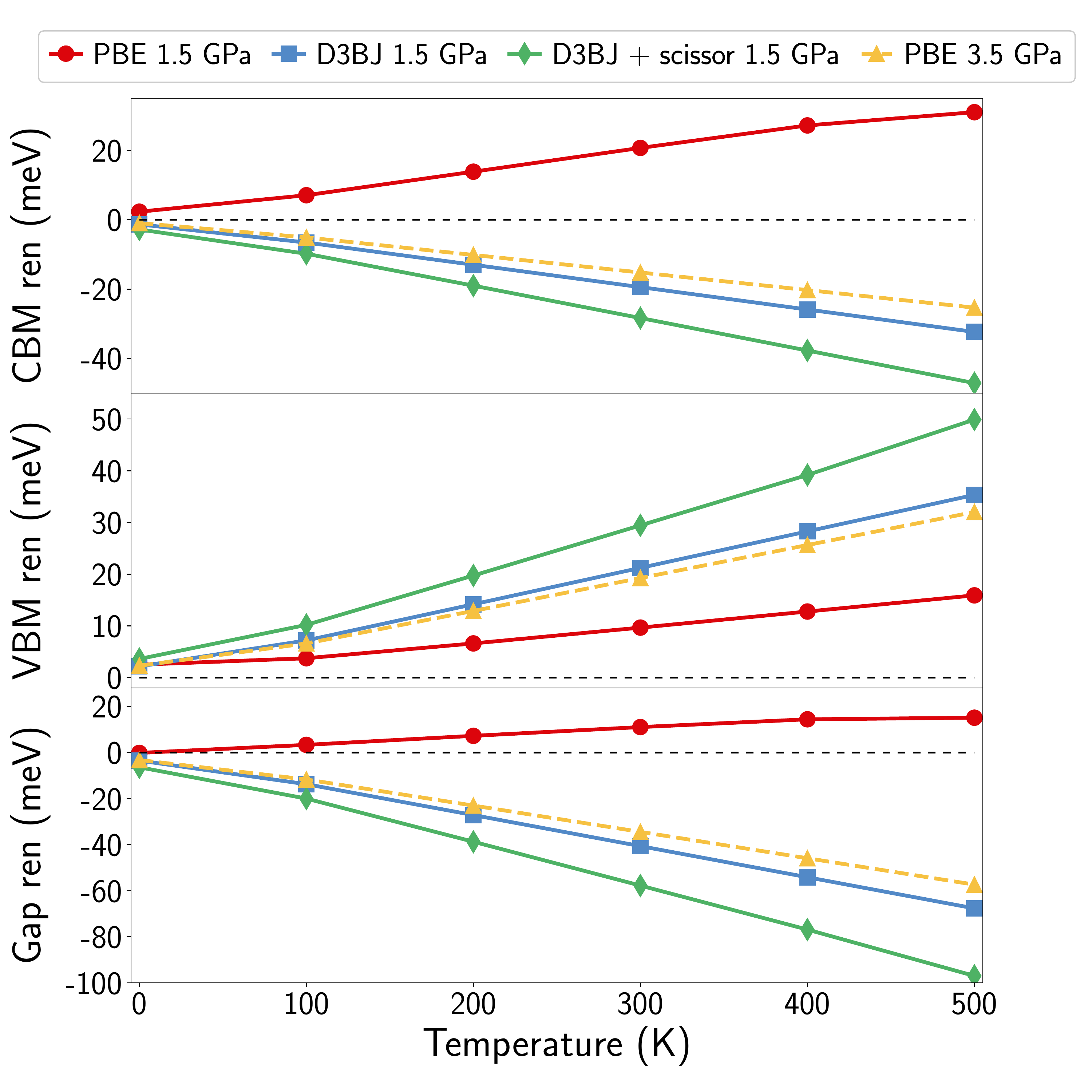}
    \caption{\textbf{Temperature-dependent renormalization of the
    band-gap edges}
    at 1.5~GPa, using DFT-D3BJ (blue squares) and PBE-GGA (red circles). The green
    curve (diamond markers) shows
    the effect of adding a scissor shift operator to DFT-D3BJ to bring
    the
    ambient pressure gap at the experimental value. Both methods
    disagree on
    the sign of the renormalization for the conduction band: it is
    positive
    for PBE-GGA and negative for DFT-D3BJ. As shown in Fig.~\ref{FigBandRenormEPI},
    this sign change occurs when the system enters the topological
    phase. Note
    that the DFT-D3BJ result at 1.5~GPa is almost identical to the
    PBE-GGA
    result at 3.5~GPa (dashed yellow line, triangle markers), that is, shifted by $P_0=-2$~GPa.}
    \label{FigVDW}
\end{figure}
The purpose of this work is to analyze the temperature dependence of 
the TPT
in BiTeI. The numerical value of the critical pressures is directly
related to
the bare band gap value, which will be corrected by electron-phonon
interaction.
From Eq.~(\ref{EqnSigmaFan}) and~(\ref{EqnSigmaDW}), we can see that the phonon
frequencies
and the bare electronic eigenvalues are key physical quantities in this
correction. This dependence goes beyond the simple electronic dispersion:
as
discussed in the main text, the band gap also plays a crucial role.
Indeed,
the denominator captures the \enquote{unlikelyhood} of couplings between
two
eigenstates with a large energy difference, rooted in perturbation
theory.
Furthermore, considering the renormalization of both band extrema
forming the
fundamental band gap, the gap energy will act as a \enquote{barrier},
giving a
lower bound on the smallest value of the denominator for couplings with
eigenstates within the subset of bands located across the gap.
Therefore, a
significant underestimation of the bare band gap will artificially
emphasize
interband couplings and bias the resulting zero-point renormalization. 
While
one could argue that this effect should be small at ${T=0}$~K, especially
for a
material with heavy atoms like BiTeI, it should not be forgotten that
it will be
amplified by the phonon occupation factor at higher temperatures, and
thus could
lead to significantly different conclusions. Lastly, since we are
dealing with
a TPT, the orbital character of the valence and conduction bands close
to their
extrema is subject to an inversion process. This phenomenon is captured
by the
wavefunction, such that a wrong band character could alter the
different matrix
elements entering the self-energy.

While DFT-D3BJ provides the most accurate descriptions of the cell
geometry and
Raman-active phonon frequencies, our results indicate that in our case,
it is
not a suitable choice to track the temperature dependence of the TPT.
On the one
hand, despite its accurate description of Van der Waals bonding between
the atomic planes, DFT-D3BJ delivers a severe underestimation of
the band gap at ambient pressure. 
This can be understood by considering the van der Waals correction as
acting
like an \enquote{effective pressure} along the $\hat{c}$~axis. As
previously
argued, such a small value of the band gap would not only flaw our
bare $P_{\text{C}}$
predictions (by the same arguments as Rusinov \emph{et
al.}~\cite{rusinov_pressure_2016}) but could also artificially
strengthen some
interband couplings in the EPI self-energy.

One could easily argue that this band gap underestimation could be
overturned
simply by applying a scissor shift operator to the eigenvalues
entering the
self-energy, or be compensated by evaluating the pressure shift that
would bring
the gap to the experimental value. We considered both these approaches.
For the
latter, we estimate that one must shift towards a negative pressure
$P_0\simeq-2$~GPa in order to bring the DFT-D3BJ gap close to the
experimental
value, thus bringing the lattice parameters very close to the PBE
relaxed values
at ambient pressure. 

About the former, we have investigated the temperature-dependent
renormalization
at 0, 1.5, 3.5 and 5~GPa, with and without the scissor shift operator.
The
results for $P=1.5$~GPa shown in Fig.~\ref{FigVDW} are quite striking:
PBE-GGA
(red circles) and DFT-D3BJ with (green diamonds) and without (blue squares) the scissor shift
disagree
about the sign of the renormalization for the CBM. This sign
disagreement is
also reflected in the total gap renormalization. 

This discrepancy brings to light a fundamental aspect of the
calculation,
namely the fact that the EPI renormalization is affected by the system's
topology. From $\mathbb{Z}_2$ topological invariant calculations,
we found that
at 1.5~GPa, the DFT-D3BJ wavefunction already is in the
\emph{topological} state
(yielding a \emph{negative} correction), while for PBE-GGA it is
still in the
\emph{trivial} state (where the correction is \emph{positive}). These are
\emph{precisely} the trends observed in Fig.~\ref{FigBandRenormEPI}. 
Moreover, if
we compare the DFT-D3BJ without scissor shift (blue squares) to the PBE-GGA
result at
3.5~GPa (dashed yellow line, triangle markers), which is roughly {$P=1.5~\text{GPa}-P_0$}, we can see that
those two
calculations, with both wavefunctions in the topological state and
with roughly
the same lattice parameters and bare band gap, deliver almost
identical results. 

Thus, even if DFT-D3BJ delivers the most accurate lattice parameters and
Raman-active phonon frequencies at 1.5~GPa when compared to experiment,
it
predicts a $\mathbb{Z}_2$ topological index that contradicts all
experimental
evidences of the phase transition. Therefore, computing the
renormalization
with DFT-D3BJ and a scissor shift operator for a given pressure would
roughly
result in applying the electron-phonon contribution from a PBE-GGA
calculation
at another pressure, shifted by $P_0$. This is especially crucial when
investigating pressures neighboring the TPT, as it can lead to wrongly
assign a
renormalization computed with a TI wavefunction to a shifted band
structure
intended to describe the trivial phase. For pressures sufficiently far
from the
transition in the TI phase, we expect the trend of a DFT-D3BJ with scissor shift
calculation to be
correct, despite a discrepancy in the numerical value of the
renormalization.
For the trivial phase, a $\mathbb{Z}_2$ calculation showed that
$0<P_{C1}<0.5$~GPa for DFT-D3BJ. Therefore, we cannot rely on such
results to
describe the ambient pressure renormalization since the resulting
wavefunction
does not predict the correct band character.


For these reasons, we chose to rely on PBE-GGA functional:
despite a clear
overestimation of the lattice parameters, it does deliver a band gap,
Rashba
splitting and Raman-active phonon frequencies that are in reasonable
overall
agreement with experiment. Moreover and most crucially, it exhibits the
right
band characters at ambient pressure and predicts a critical pressure
$P_{\text{C1}}$ in agreement with experimental evidence.

\section{Pressure and temperature-dependent Rashba Splitting}\label{AppendixRashba}

\begin{figure*}
    \centering
    \includegraphics[width=\linewidth]{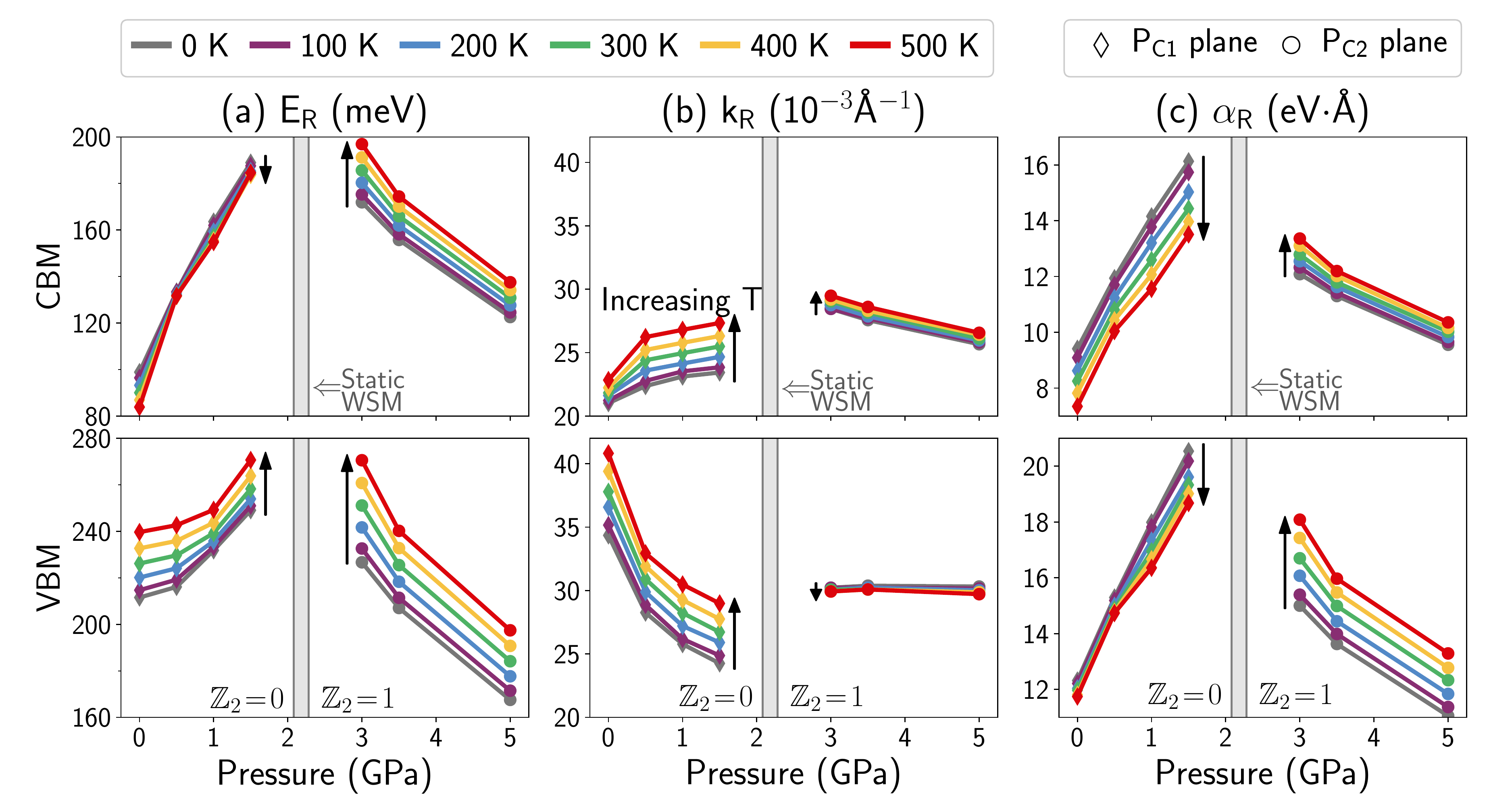}
    \caption{\textbf{Pressure and temperature-dependent Rashba parameters}, for conduction (top panels) and valence bands (bottom panels), near the minimum band gap of the $P_{\text{C1}}$ (diamond markers) and $P_{\text{C2}}$ (round markers) planes, as defined in Fig.~\ref{FigStructure}(b). Note that for $P<P_{\text{C1}}$ these quantities were computed in {the H-A} direction in the $P_{\text{C1}}$ plane, hence the difference with the data in Table~\ref{TableRashba} for the conduction band at ambient pressure, where it was computed in the {A-L} direction for comparison with experimental data. For $P>P_{\text{C2}}$ it was computed in the {A-L} direction in the $P_{\text{C2}}$ plane.}
    \label{FigRashbaPT}
\end{figure*}
As discussed in the main text, one of the assumptions underlying our work is that the Weyl nodes trajectory throughout the WSM phase is not qualitatively affected by temperature. In other words, we suppose that the $k_z$ location of the $P_{\text{C1}}$ and $P_{\text{C2}}$ planes remains constant; hence we only investigated EPI and TE within these planes. Without having access to the pressure and temperature dependence of the band structure in the full BZ, we can nevertheless gain significant insight by tracking the pressure and temperature dependence of the different parameters characterizing Rashba splitting within these planes. The Rashba energy, $E_\text{R}$, captures the steepness of the band's slope in the vicinity of the band gap. The Rashba momentum, $k_\text{R}$, tracks the distance between the band extrema and the degeneracy point at $(0,0, k_z)$, that is, the temperature-dependent band gap location within a given $k_z$ plane. Finally, the Rashba parameter, $\alpha_\text{R}$,
 captures the strength of the Rashba interaction (Eq.~(\ref{EqnRashba})), which is related to the potential gradient along the $\hat{c}$~axis~\cite{2008Spis}.

Fig.~\ref{FigRashbaPT} presents the results for the Rashba energy~(a), Rashba momentum~(b) and Rashba parameter~(c), for both the valence (bottom panels) and conduction bands (upper panels). We emphasize that these parameters were computed for the minimal band gap, namely on the ({H-A},~$P_{\text{C1}}$) line for the trivial phase and on the \mbox{(A-L,~$P_{\text{C2}}$)} for the TI phase. Hence, the numerical values for ambient pressure in this figure should not be compared to experimental data, as available ARPES measurements were done on the \mbox{(A-L,~$P_{\text{C1}}$)} line. The trends reported here nevertheless agree with those of Monserrat and Vanderbilt~\cite{monserrat_temperature_2017} for the \mbox{(A-L, $k_z=\frac{\pi}{c})$} direction.

As can be seen in Fig.~\ref{FigRashbaPT}, pressure has a stronger effect than temperature on the Rashba parameters. This behavior can be qualitatively understood as increasing temperature enhances atomic vibrations around their slightly shifted equilibrium positions, while pressure modifies more significantly these equilibrium positions by reducing the distance between the atomic planes. Hence, for a given temperature, increasing pressure enhances the potential gradient along the $\hat{c}$ axis, thus increasing $\alpha_{\text{R}}$. On the contrary, for a given pressure, temperature increases the complex interplay between the electrons and the lattice, yielding a reduction of $\alpha_\text{R}$. These are the behaviors observed in the trivial phase [Fig.~\ref{FigRashbaPT}(c), diamond markers].

The TI phase displays the opposite behavior [Fig.~\ref{FigRashbaPT}(c), round markers]: $\alpha_{\text{R}}$ is reduced by pressure and increased by temperature. To understand this behavior, we must analyze more carefully both the Rashba energy and momentum, as $\alpha_\text{R}$ captures the interplay between those two quantities. On the one hand, for all pressures, the Rashba energy [Fig.~\ref{FigRashbaPT}(a)] almost steadily increases with temperature, except for the CB in the trivial phase, which shows a very weak decrease with temperature. On the other hand, the Rashba momentum is almost unaffected by pressure and temperature in the TI phase [Fig.~\ref{FigRashbaPT}(b), round markers], while it increases almost steadily with temperature for pressures below the TPT. 

The opposite trends of  $\alpha_{\text{R}}$ between the trivial and TI phase can, therefore, be understood by two observations. On the one hand, in both phases, the pressure dependence of $\alpha_{\text{R}}$ is governed by the Rashba energy, which increases in the trivial phase and decreases in the TI phase. On the other, their opposite temperature dependence is caused by the Rashba momentum. In the trivial phase, both $E_{\text{R}}$ and $k_{\text{R}}$ increase with temperature, the latter being more significant than the former, hence the decrease of $\alpha_{\text{R}}$. In the TI phase, the increase of $\alpha_{\text{R}}$ captures the change of the Rashba energy, as $k_{\text{R}}$ is scarcely affected by temperature. We cannot, however, determine if this behavior is linked to the TPT or merely a consequence of an increased bonding along $\hat{c}$ induced by increased pressure. This question could be addressed by investigating the temperature dependence of the high-pressure Rashba splitting in other bismuth tellurohalides, BiTeBr and BiTeCl, which are not known to exhibit a pressure-induced TPT~\cite{rusinov_pressure_2016}.

Lastly, we verified that the trends presented in Fig.~\ref{FigRashbaPT} for the TI phase do not vary if the temperature-dependent band structure is evaluated in the $P_\text{C1}$ plane rather than the $P_\text{C2}$ plane, remaining in the {A-L} direction. Hence, it seems reasonable that these trends, including the quasi temperature-independence of the Rashba momentum,  could be applied to any $k_z$ between those planes. We also verified that the temperature-dependent values of $P_\text{C2}$ are not significantly altered if considering the EPI and TE corrections computed in the $P_\text{C1}$ plane.

\section{Temperature-dependent Weyl semimetal phase
width}\label{AppendixWSM}

\begin{table*}
\setlength\extrarowheight{1pt}
\caption{\textbf{Temperature-dependent WSM phase
boundaries,} obtained with the extrapolation
procedure described in Sect.~\ref{WSMResults}.
$P_{\text{C1}}$ marks the phase transition from a trivial insulator
to a WSM, while $P_{\text{C2}}$ marks the transition from WSM to
$\mathbb{Z}_2$ topological insulator. The EPI+TE values are plotted
in Fig.~\ref{FigPTDiagram} 
(black markers). The middle and left
sections display the critical pressures obtained when considering only
the EPI or TE contribution to the renormalization.}\label{TableWSM}
\begin{tabular}{|c||c|c|c||c|c|c||c|c|c|}
\hline\hline
Temperature&\multicolumn{3}{c||}{EPI only}&\multicolumn{3}{c||}{TE
only}&\multicolumn{3}{c|}{EPI+TE}\\
\cline{2-10}
(K) & $P_{\text{C1}}$  \footnotesize{(GPa)} & $P_{\text{C2}}$ 
\footnotesize{(GPa)} & width  \footnotesize{(GPa)}& $P_{\text{C1}}$
\footnotesize{(GPa)}& $P_{\text{C2}}$ \footnotesize{(GPa)} & width 
\footnotesize{(GPa)}& $P_{\text{C1}}$ \footnotesize{(GPa)}&
$P_{\text{C2}}$  \footnotesize{(GPa)}& width  (GPa)\\[3pt]
\hline
 Static & 2.08 & 2.28 & 0.20& 2.08 & 2.28 & 0.20& 2.08 & 2.28 & 0.20\\
0   & 2.08 & 2.31 & 0.23 &2.08  & 2.28 & 0.20 & 2.08 &2.31  &0.23  \\
100 & 2.10 & 2.42 & 0.32 & 2.09 &2.39  &0.30  & 2.14 &2.53  &0.39  \\
200 & 2.14 & 2.57 & 0.44 & 2.16 & 2.55 & 0.39 & 2.24 & 2.82 &0.58  \\
300 & 2.17 & 2.73 & 0.56 & 2.24 & 2.70 & 0.46 & 2.30 & 3.09 & 0.79 \\
400 & 2.20 & 2.89 & 0.69 & 2.33 & 2.83 & 0.50 & 2.32 & 3.33 & 1.01 \\
500 & 2.22 & 3.05 & 0.83 &2.46  & 2.95& 0.51 & 2.30 & 3.55 & 1.25 \\
\hline\hline
\end{tabular}
\end{table*}

In Table~\ref{TableWSM}, we report the numerical values for the
temperature-dependent WSM phase boundaries, namely $P_{\text{C1}}(T)$
and $P_{\text{C2}}(T)$. These values were extracted from Fig.~\ref{FigGapPT} 
using the extrapolation procedure described in
Sect.~\ref{WSMResults}
and are explicitly plotted in Fig.~\ref{FigPTDiagram}. 
For comparison, we also include the
temperature-dependent critical pressures obtained by considering
the EPI and TE corrections independently. 



%
\end{document}